\documentclass[twocolumn,floatfix,prb,aps,showpacs]{revtex4-2}
\usepackage{graphicx,amsmath,amssymb,color,nicefrac,multirow, makecell, ragged2e,float}
\usepackage[unicode=true,colorlinks=true,linkcolor=blue,citecolor=blue,urlcolor=blue]{hyperref}
\usepackage[titletoc,title]{appendix}

\newcommand{\be}{\begin{equation}}
\newcommand{\ee}{\end{equation}}

\newcommand{\ba}{\begin{eqnarray}}
\newcommand{\ea}{\end{eqnarray}}

\newcommand{\IO}{\mathcal{O}}
\newcommand{\IH}{\mathcal{H}}
\renewcommand{\vec}[1]{\mbox{\boldmath$#1$}}

\def\beq{\begin{eqnarray}}
\def\eeq{\end{eqnarray}}

\makeatletter
\newcommand*{\rom}[1]{\expandafter\@slowromancap\romannumeral #1@}
\makeatother
\begin{document}
\title{Anderson localization in fractional quantum Hall effect}
\author{Songyang Pu$^1$, G. J. Sreejith$^2$, J. K. Jain$^1$}
\affiliation{$^1$Department of Physics, 104 Davey Lab, Pennsylvania State University, University Park, Pennsylvania 16802,USA}
\affiliation{$^2$Indian Institute of Science Education and Research, Pune 411008, India}
\date{\today}
\begin{abstract} 
The interplay between interaction and disorder-induced localization is of fundamental interest. This article addresses localization physics in the 
fractional quantum Hall state, where 
both interaction and disorder have nonperturbative consequences. We provide compelling theoretical evidence that the localization of a single quasiparticle of the fractional quantum Hall state at filling factor $\nu=n/(2n+1)$  has a striking {\it quantitative} correspondence to the localization of a single electron in the $(n+1)$th Landau level. By analogy to the dramatic experimental manifestations of Anderson localization in integer quantum Hall effect, this leads to predictions in the fractional quantum Hall regime regarding the existence of extended states at a critical energy, and the nature of the divergence of the localization length as this energy is approached. Within a mean field approximation these results can be extended to situations where a finite density of quasiparticles is present.
\end{abstract}
\maketitle

{\it Introduction:}
The scaling theory of localization~\cite{Abrahams79} made the remarkable prediction that arbitrary weak random disorder localizes all eigenstates of a single electron in two dimensions (2D), implying an absence of a metallic phase for non-interacting electrons in 2D.  The physics of localization also played a central role in Laughlin's explanation~\cite{Laughlin81} of the origin of plateaus in the integer quantum Hall effect (IQHE)~\cite{Klitzing80}, leading to extensive experimental and theoretical investigation of Anderson localization of a single electron in the presence of a magnetic field.
The observation of IQHE implies, unlike at zero magnetic field, existence of extended single particle states in the presence of a magnetic field. It has been argued that, in the asymptotic limit, extended states occur at a single critical energy $E_c$ in each Landau level (LL). Experimental measurements of the temperature dependence of the width of the transition region from one plateau to next suggest a power law divergence of the localization length as the critical energy is approached~\cite{Li05,Li09}, although slightly dissimilar exponents have been observed in different experiments and for different transitions. Many theoretical studies have attempted to determine the value of the critical exponent characterizing the divergence of the localization length~\cite{DasSarma96,Huckestein95,Chalker88,Arovas88,Zirnbauer94,Bhatt21,Zhu21b}. We note that recent work has called the notion of scaling into question and suggested that the exponent observed in numerical calculations and experiments is only an 
``effective" exponent which is not universal but model dependent~\cite{Zirnbauer19,Dresselhaus21}.

Interaction between particles significantly complicates the problem. For example, interaction is thought to be responsible for a metal-insulator transition in 2D electron systems at zero magnetic field~\cite{Kravchenko03}. 
The objective of this work is to address the localization physics in the fractional quantum Hall effect (FQHE) regime, where both the interaction and disorder cause highly nontrivial, non-perturbative phenomenology. 
From a microscopic perspective, beginning with interacting electrons in a disordered potential is not practical or fruitful, and exact diagonalization studies are inadequate because they are limited to very small systems, whereas, as we see below, it is necessary to go to rather large systems to capture the thermodynamic behavior. Fortunately, the composite fermion (CF) theory~\cite{Jain89,Jain90,Jain07} opens a new avenue for tackling this problem. According to the CF theory, the non-perturbative role of interaction is to create composite fermions, and the problem of strongly interacting electrons in the FQHE regime maps into that of weakly interacting composite fermions in the IQHE regime. This suggests the possibility that the localization physics of the FQHE is analogous to that of the IQHE~\cite{Jain90b,Kivelson92}, but a microscopic confirmation of such a correspondence has been lacking.  That question has motivated the present work. The primary conclusion of our work is to demonstrate that in the presence of a disorder potential, composite fermions in the $n$th CF-LL [called $\Lambda$ level ($\Lambda$L)] behave, to a surprising degree, as electrons in the $n$th LL, implying that the localization physics in the FQHE corresponds to that in the IQHE in its universal {\it as well as} nonuniversal aspects.

{\it Model:} We begin by considering a single quasiparticle (qp) or quasihole (qh) in the presence of an 
disorder potential $H=\sum_{\alpha,k} \epsilon_k \delta(\vec{r}_\alpha-\vec{w}_k)$, which represents a random distribution of short-range impurities
at positions $\{\vec{w}_k\}$ with random on-site energies $\epsilon_k$.  We will assume that disorder strength is weak compared to the FQHE gap, in the sense that it neither creates quasiparticles or quasiholes out of the FQHE vacuum nor significantly distorts the wave function of the already present quasiparticles or quasiholes. The objective then reduces to diagonalizing the above problem in the basis $\{ \Psi_{w_j} \}$, where $\Psi_{w_j}$ is the wave function of the quasiparticle or quasihole 
localized at $\vec{w}_j$. Remembering that the basis is not orthogonal, the eigenfunctions $\Phi=\sum_i c_i \Psi_{w_i}$ and the eigenenergies $E$ satisfying $H\Phi=E\Phi$ are given by the solutions of the matrix equation $O^{-1}H c=E c$, with $O_{ij}=\langle \Psi_{w_i} | \Psi_{w_j}\rangle$, $H_{ij}=\langle \Psi_{w_i} |H| \Psi_{w_j}\rangle$, and $c=(c_1, \cdots, c_N)^{\rm T}$. The quantity $H_{ij}$ is the tunneling amplitude for $i\neq j$. Note that $\epsilon_j$ can be positive or negative; we keep only the $\Psi_{w_i}$ localized at  the ``attractive" impurities in our basis, although all impurities contribute to $H_{ij}$. We will assume that the number of impurities is smaller than the total number of available orbitals, ensuring that the $\Psi_{w_j}$ in our basis are linearly independent. $H_{ij}$ and $O_{ij}$ are in general complex due to the breaking of time reversal invariance by the magnetic field, which is fundamentally responsible for the strikingly different behavior in a magnetic field.

The quantities $H_{ij}$ and $O_{ij}$ can be evaluated (numerically) for the FQHE provided we know $\Psi_{w_j}$. One can imagine resorting to exact diagonalization studies but those are restricted to systems that are too small to be meaningful for the issue at hand, given that even a single localized quasiparticle or quasihole has a rather large size. We instead use the CF theory, which has clarified that, microscopically, the quasiparticles are composite fermions in an almost empty $\Lambda$L, and quasiholes are missing composite fermions in an almost full $\Lambda$L~\cite{Jain89,Jain89b,Jain07,Jeon03b,Jeon04}. To construct their wave function, we recall that the wave function of a single electron localized at $w=w_x+iw_y$ in the $n$th LL at effective magnetic field $B^*$ is given by 
\be
\phi^{(n)}_w(z,B^*)=\sum_m [\eta^{(n)}_{m}(w,B^*)]^* \eta^{(n)}_{m}(z,B^*)
\ee
where the particle coordinate is defined as $z=x+iy$, 
and $m$ is the angular momentum index. The single particle angular momentum orbitals in the $n$th LL are given by 
\be
\eta^{(n)}_{m}(z,B^*)={1\over {\ell^{*}}}\sqrt{n!\over 2\pi 2^m(m+n)!} \frac{z^m}{\ell^{*m}}L_n^m\left({|z|^2\over {2\ell^{*2}}}\right) e^{-{|z|^2\over 4\ell^{*2}}}
\ee
with the effective magnetic length $\ell^*=\sqrt{\hbar c/eB^*}$ and $L_n^m(t)$ is the associated Laguerre polynomial in Rodrigues definition.
The wave function $\Psi_w^{{\rm qp}-n}$ for the CF quasiparticle of  
the $\nu=n/(2pn+1)$ state, located at $w$, is constructed by composite-fermionizing the state with $n$ filled LLs and one additional electron localized at $w$ in the $(n+1)$th LL.  
To give an explicit example, the wave function of a quasiparticle of the $\nu=2/(4p+1)$ state 
is given 
by
\begin{eqnarray}
\Psi^{\rm qp-2}_{w} &=& {\cal P}_{\rm LLL} 
\left|\begin{array}{ccc}
\phi^{(2)}_{w}(z_1,B^*) & \phi^{(2)}_{w}(z_2,B^*) &\ldots\\
\eta^{(1)}_{-1}(z_1,B^*) & \eta^{(2)}_{-1}(z_2,B^*) &\ldots\\
\eta^{(1)}_{0}(z_1,B^*) & \eta^{(2)}_{0}(z_2,B^*) &\ldots\\
\vdots&\vdots&\ldots \\
\eta^{(1)}_{N_1-2}(z_1,B^*) & \eta^{(2)}_{N_1-2}(z_2,B^*) &\ldots\\
\eta^{(0)}_{0}(z_1,B^*) & \eta^{(0)}_{0}(z_2,B^*) &\ldots\\
\eta^{(0)}_{1}(z_1,B^*) & \eta^{(0)}_{1}(z_2,B^*) &\ldots\\
\vdots&\vdots&\ldots \\
\eta^{(0)}_{N_0-1}(z_1,B^*) & \eta^{(0)}_{N_0-1}(z_2,B^*) &\ldots\\
\end{array}
\right| \nonumber \\
& & \;\;\;\;\times  \left[\prod_{i<k=1}^N(z_i-z_k) e^{-\sum_j|z_j|^2/4\ell_1^2}\right]^{2p}\;.
\label{CFQP}
\end{eqnarray}
Here, $\ell_1$ is the magnetic length at $\nu=1$, and $N_0$ and $N_1$ are number of composite fermions in the lowest two $\Lambda$Ls (we set $n=0$ for the lowest $\Lambda$ level), and the total number of particles is $N=N_0+N_1+1$. The symbol ${\cal P}_{\rm LLL}$ represents projection into the lowest LL (LLL), which we evaluate using the standard Jain-Kamilla method~\cite{Jain97,Jain97b}. 
The relation 
$\ell^{*-2}+2p\ell_1^{-2}=\ell^{-2}$
ensures that the product wave function has the gaussian factor corresponding to the actual external magnetic field. 
The wave functions for other quasiparticles and quasiholes (see Appendix~\ref{additionalQPQH}) can be constructed analogously. These wave functions have been tested against exact wave functions and found to be extremely accurate representations of the exact Coulomb wave functions~\cite{Jain07}. As a result, the conclusions derived below from these wave functions can be considered to be reliable. 

The key question is if the above FQHE problem is equivalent, modulo a rescaling of parameters, to an IQHE problem, for which $H_{ij}$ and $O_{ij}$ can be obtained analytically. Such a correspondence, should it exist, would be powerful, because it allows us to carry our knowledge about the localization physics of the IQHE over to the FQHE.  We demonstrate below a close correspondence between 
\be
H(w_j,\epsilon_j,B) \iff \IH(w_j,\epsilon^*_j=\epsilon_j B^*/B,B^*).
\ee
\be
O(w_j,\epsilon_j,B) \iff \IO(w_j,\epsilon^*_j=\epsilon_j B^*/B,B^*).
\ee
We use $H$ and $O$ for the matrix elements of FQHE, while $\IH$ and $\IO$ for the matrix elements of IQHE. We note that the sample size and the impurity positions remain identical in the FQHE and the IQHE systems in laboratory units; if the impurity positions are measured in units of the magnetic length, then we must scale them by a factor $(\ell^*/\ell)$.

{\it Results:} 
The natural basis set in our problem, consisting of states with the quasiparticle localized at different impurity positions, is not orthogonal. Apart from relating the matrix elements of the Hamiltonian, establishing a correspondence between the single quasiparticle localization problems in IQHE and FQHE also entails relating the overlap matrix of the basis states.

\begin{figure}[t]
	\includegraphics[width=0.48\columnwidth]{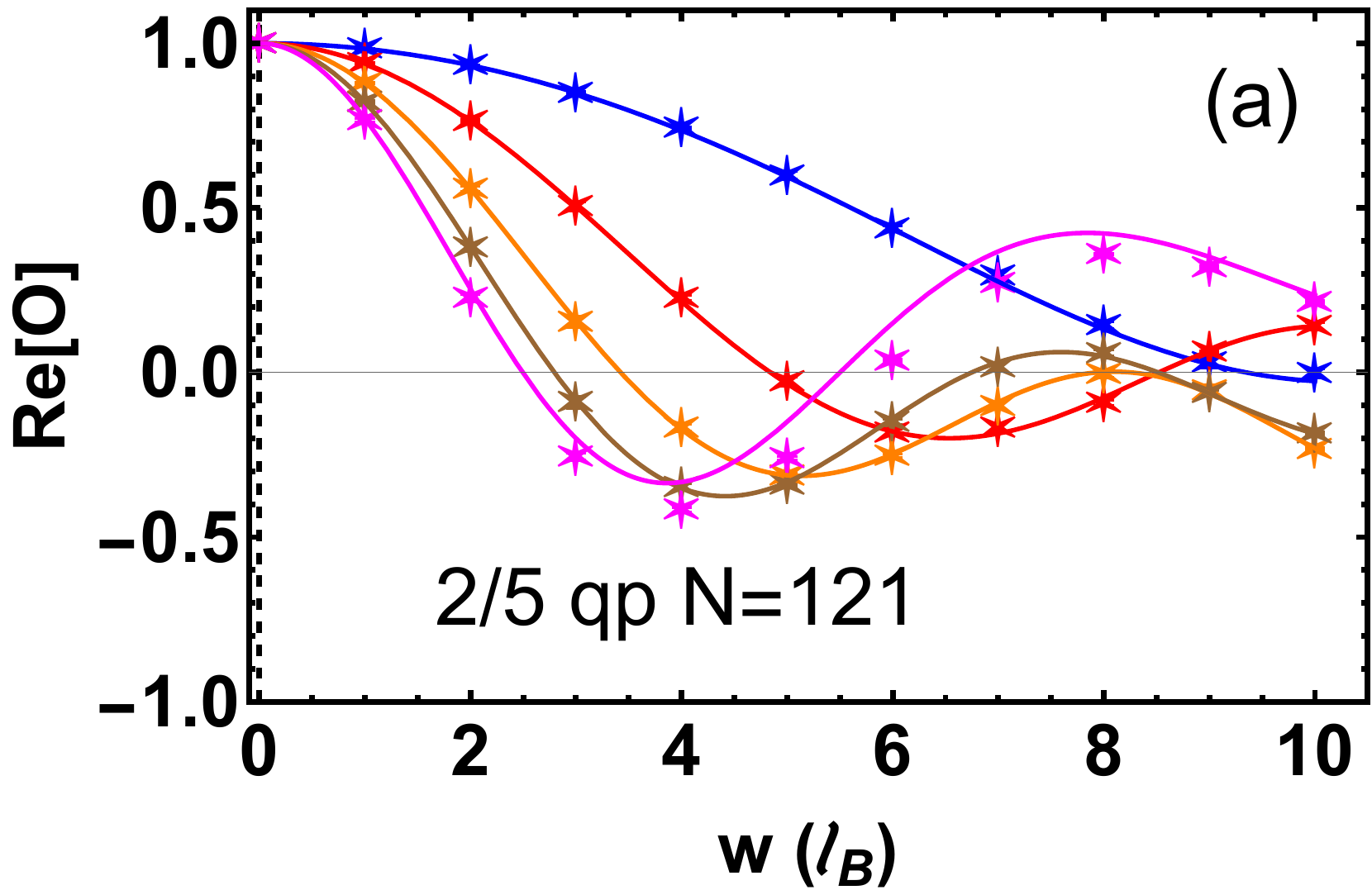} 
	\includegraphics[width=0.48\columnwidth]{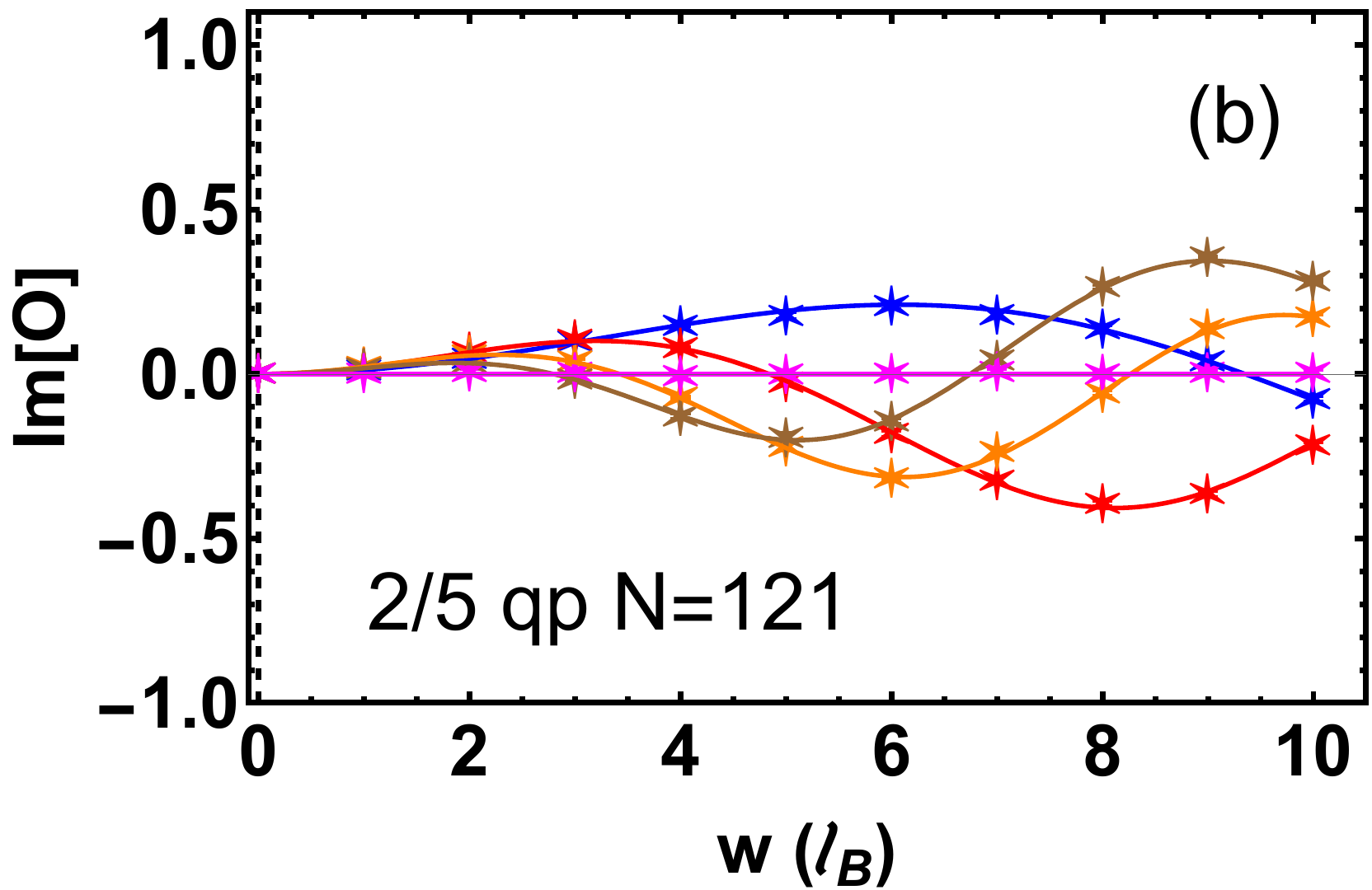} 
    \includegraphics[width=0.48\columnwidth]{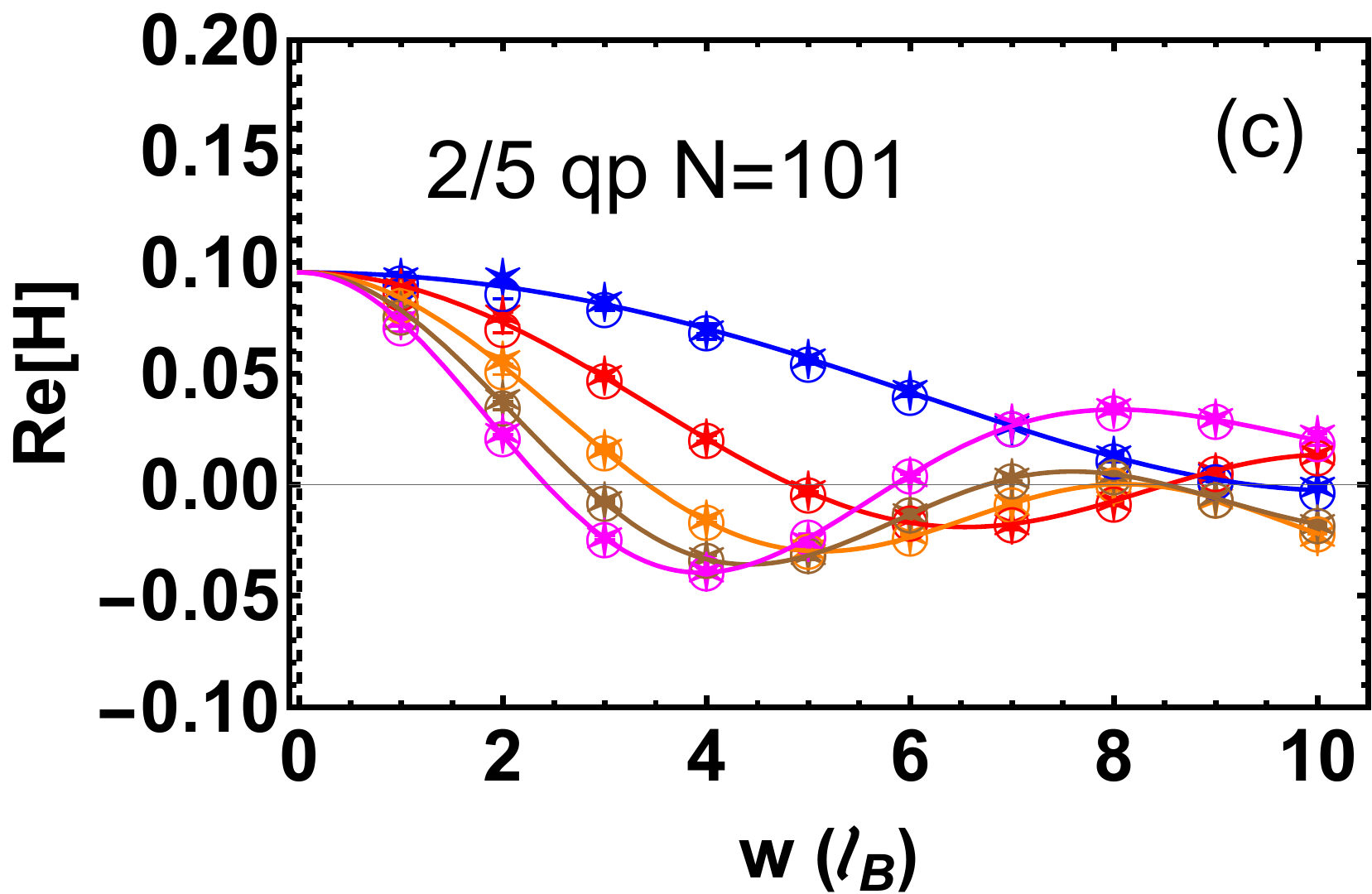} 
	\includegraphics[width=0.48\columnwidth]{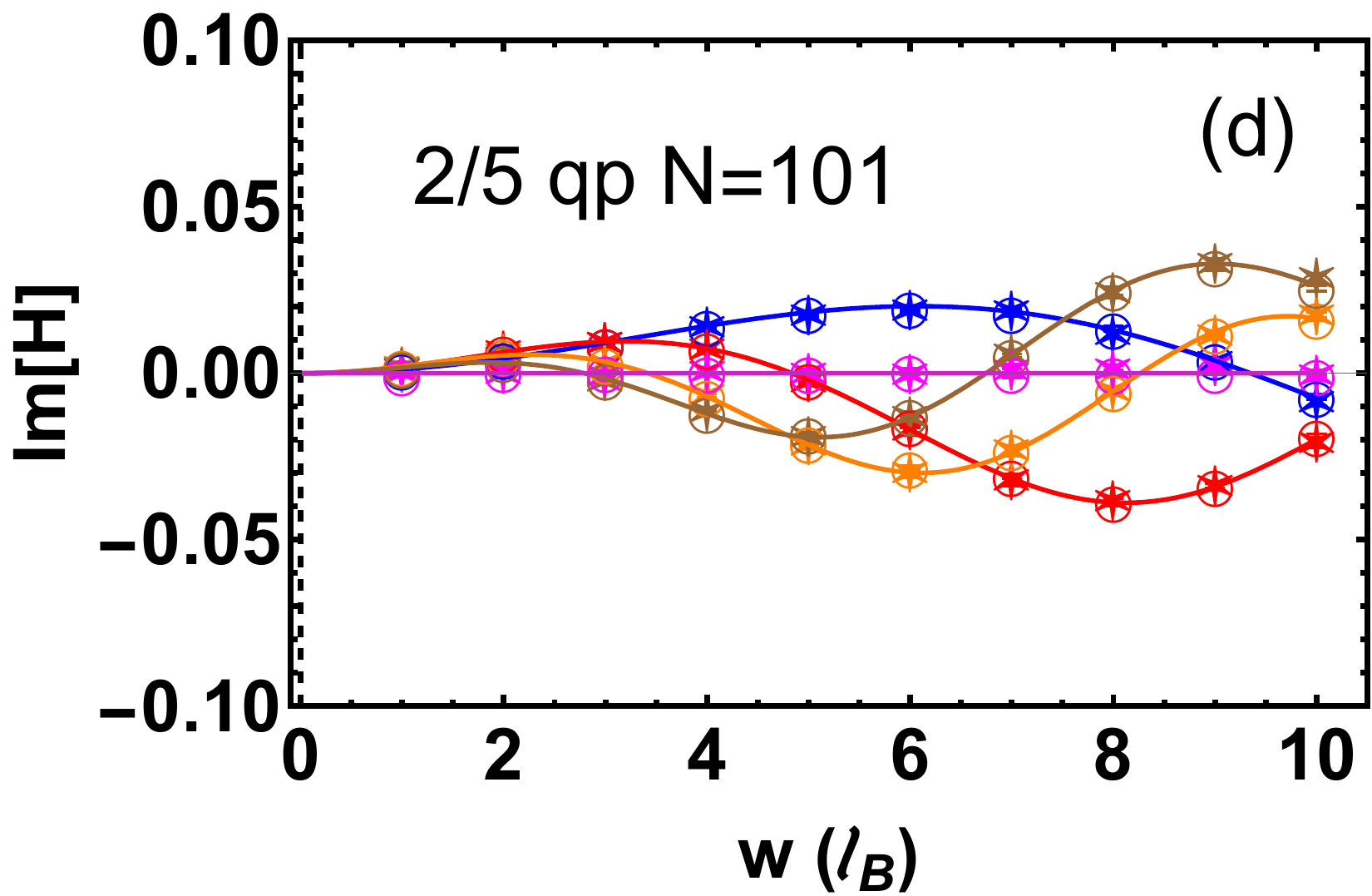} 
    \includegraphics[width=0.48\columnwidth]{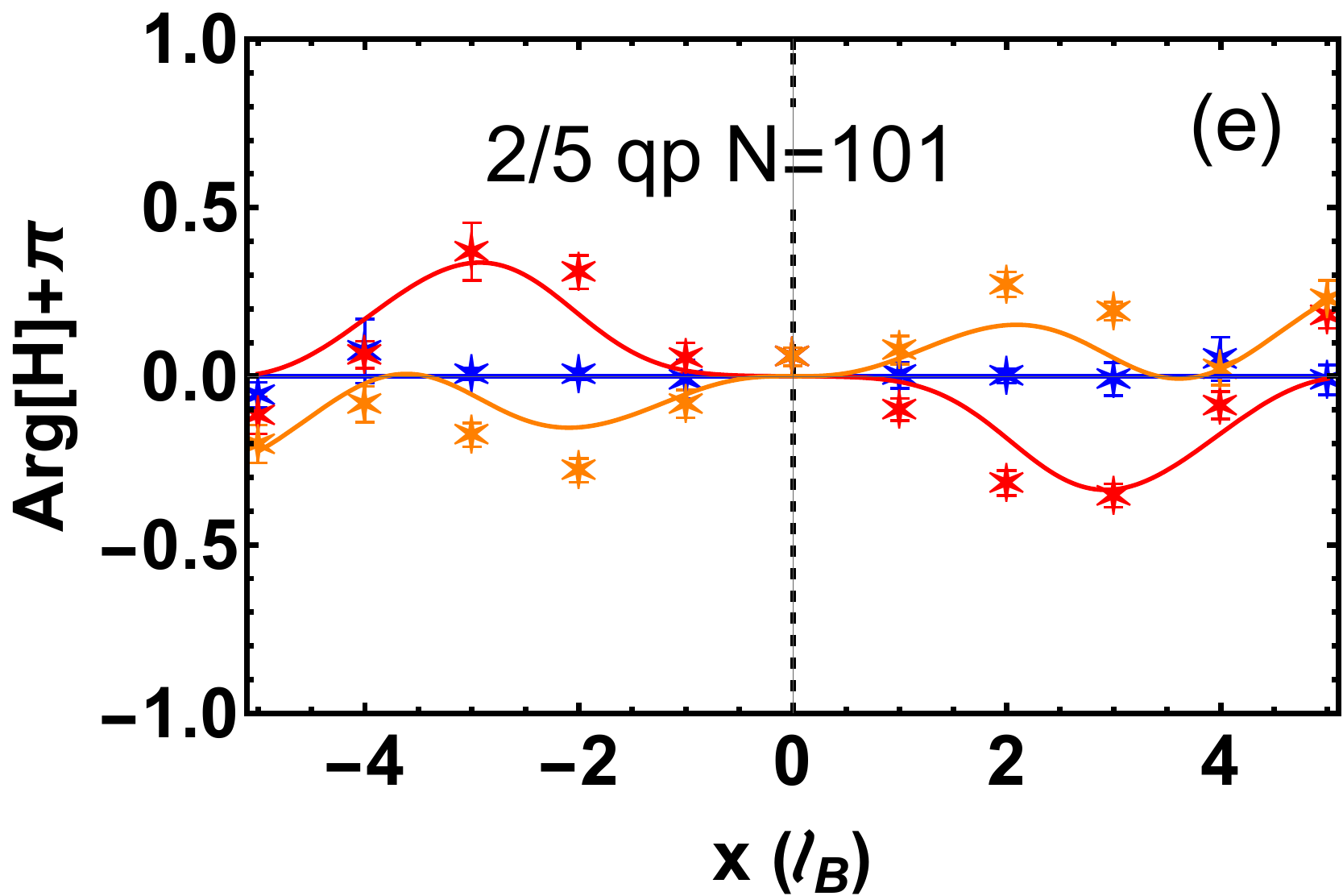} 
	\includegraphics[width=0.48\columnwidth]{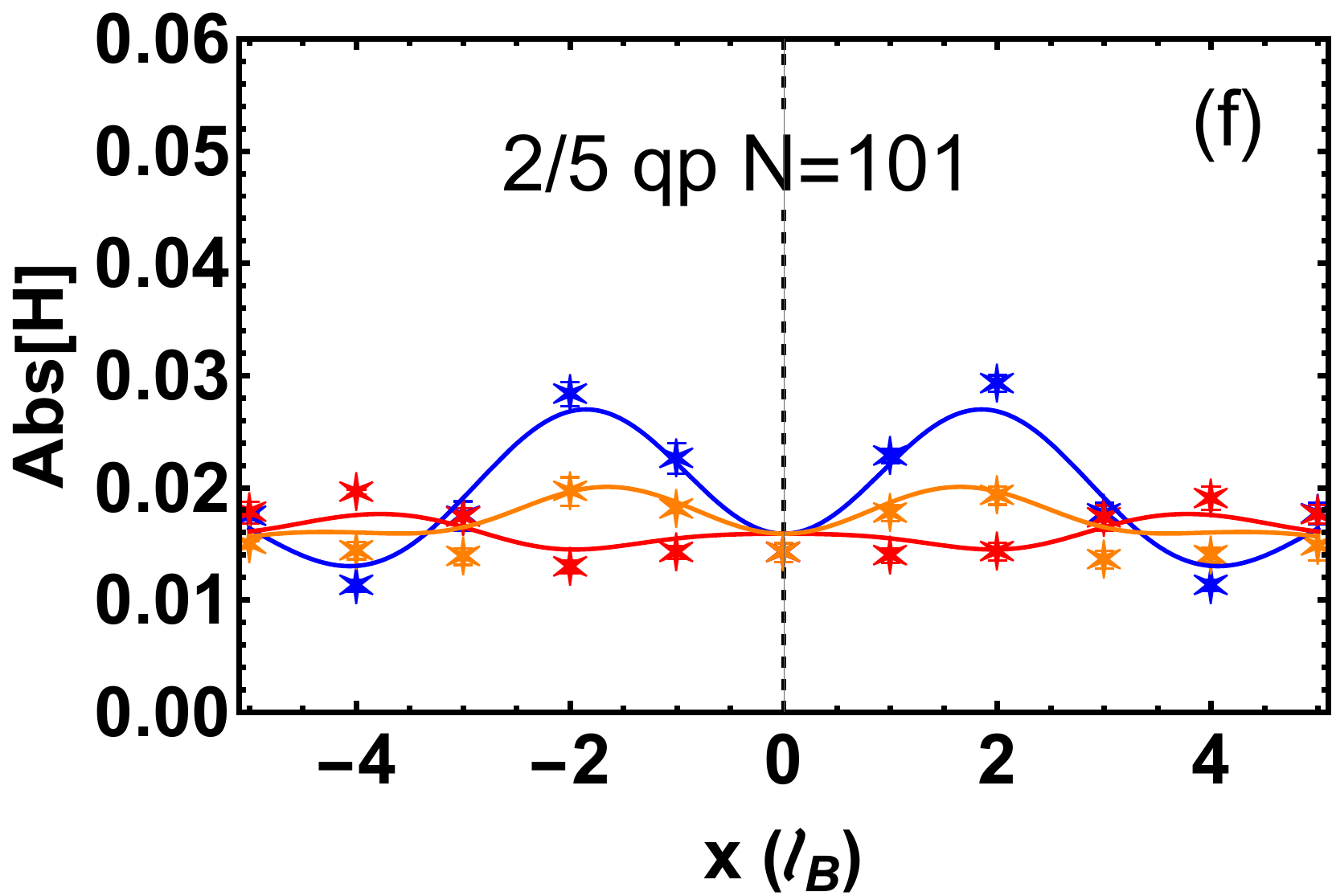} 
   \includegraphics[width=0.48\columnwidth]{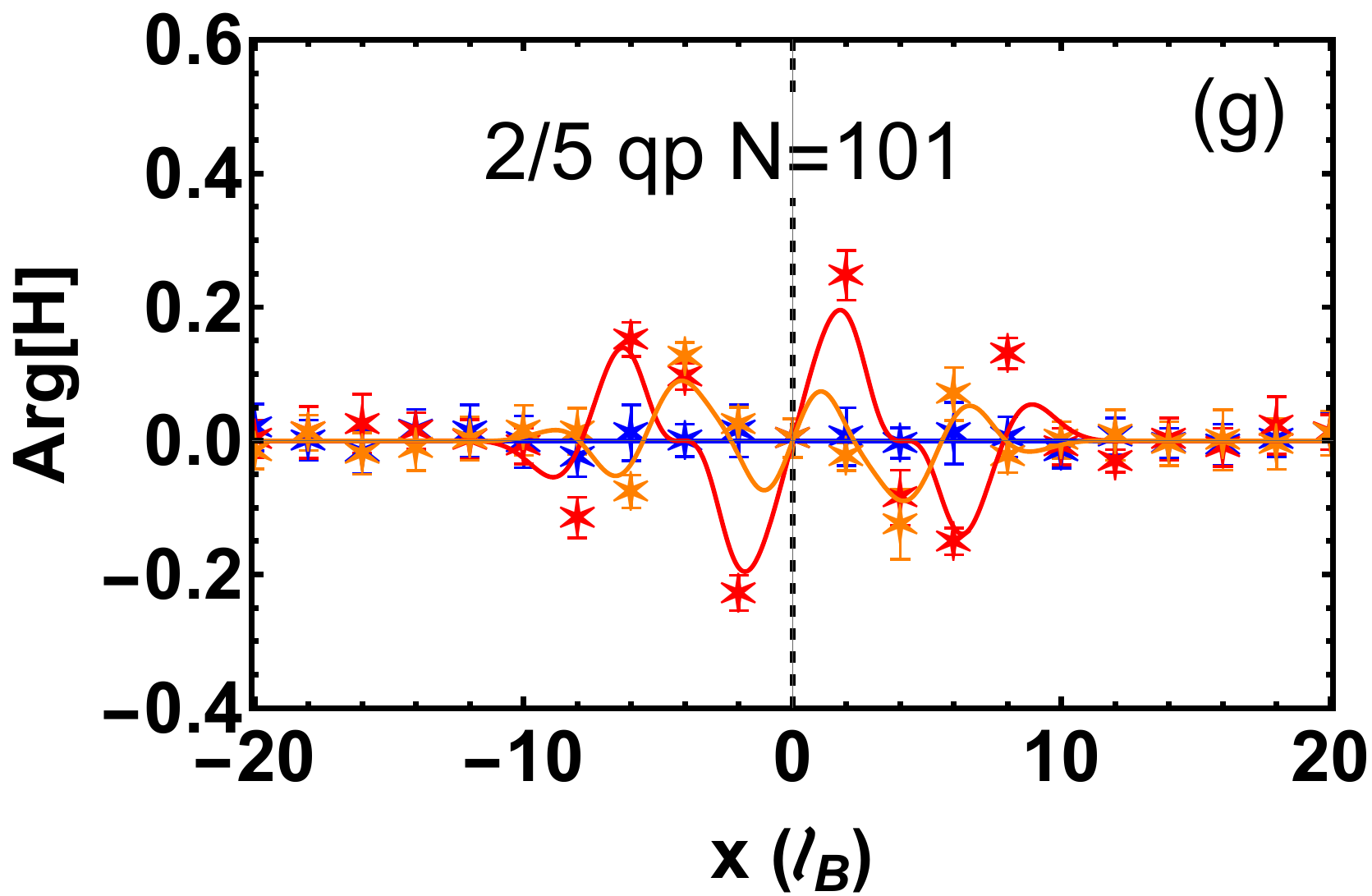} 
	\includegraphics[width=0.48\columnwidth]{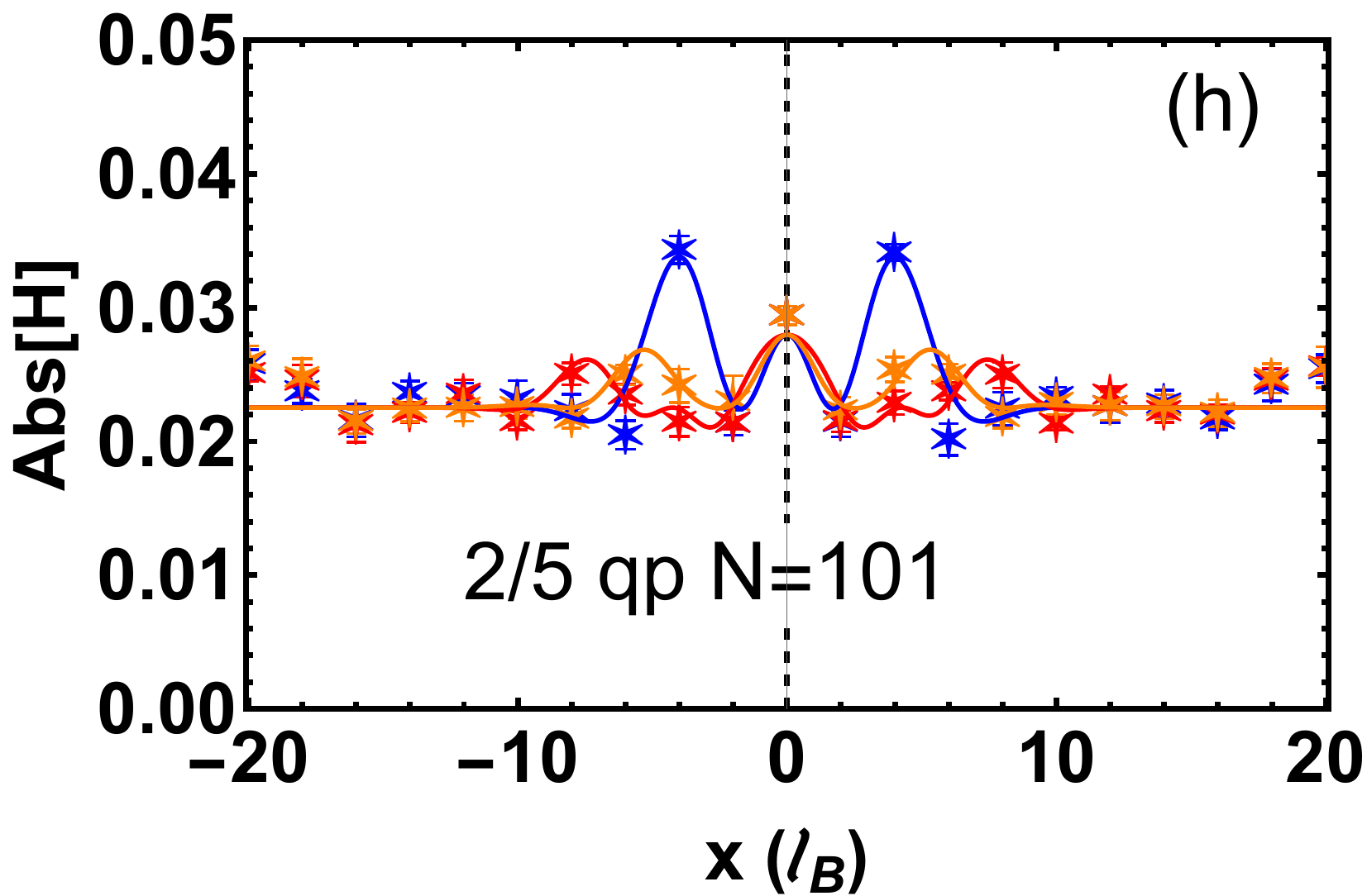} 
	\caption{This panel show various matrix elements for the quasiparticle of the 
	$\nu=2/5$ state (stars) and compares them with the corresponding analytical results for an electron in the third Landau level at an effective magnetic field (solid lines). (We stress that the solid lines have no fitting parameters in this and subsequent figures.) Panels (a) and (b) show the real part and imaginary part of the overlap matrix elements $\langle \Psi^{{\rm qp}-2}_{{w\over 2}e^{i\theta}}|\Psi^{{\rm qp}-2}_{w\over 2}\rangle$, for $\theta=\pi/6$ (blue); $\theta=\pi/3$ (red); $\theta=\pi/2$ (orange); $\theta=\pi/2$ (brown); and $\theta=\pi$ (magenta).  Panels (c) and (d) show the real part and imaginary part of the tunneling matrix elements $\langle \Psi^{{\rm qp}-2}_{{w\over 2}e^{i\theta}}|\sum_i \delta\left(\vec{r}_i-{w\over 2}e^{i\theta}\right)|\Psi^{{\rm qp}-2}_{w\over 2}\rangle$ with the same color code. (We stress that the solid lines are not fits.)  Panel (e) - (h) show the phase and the modulus of the impurity assisted tunneling matrix element $\langle\Psi^{{\rm qp}-2}_{-{w\over 2}}|\sum_i \delta\left(\vec{r}_i-xe^{i\theta'}\right)|\Psi^{{\rm qp}-2}_{w\over 2}\rangle$, for $\theta'=0$ (blue);  $\theta'=\pi/4$ (orange); and $\theta'=\pi/2$ (red). Panel (e) and (f) correspond to $w=5$, and panel (g) and (h) correspond to $w=8$. All length are quoted in units of the magnetic length at $\nu=2/5$. The number of particles $N$ in the Monte Carlo calculation is shown on each panel; the results represent the thermodynamic limit. We take $N_0=N_1$.
 }
	\label{O25}
\end{figure}

\begin{figure}[t]
	\includegraphics[width=0.48\columnwidth]{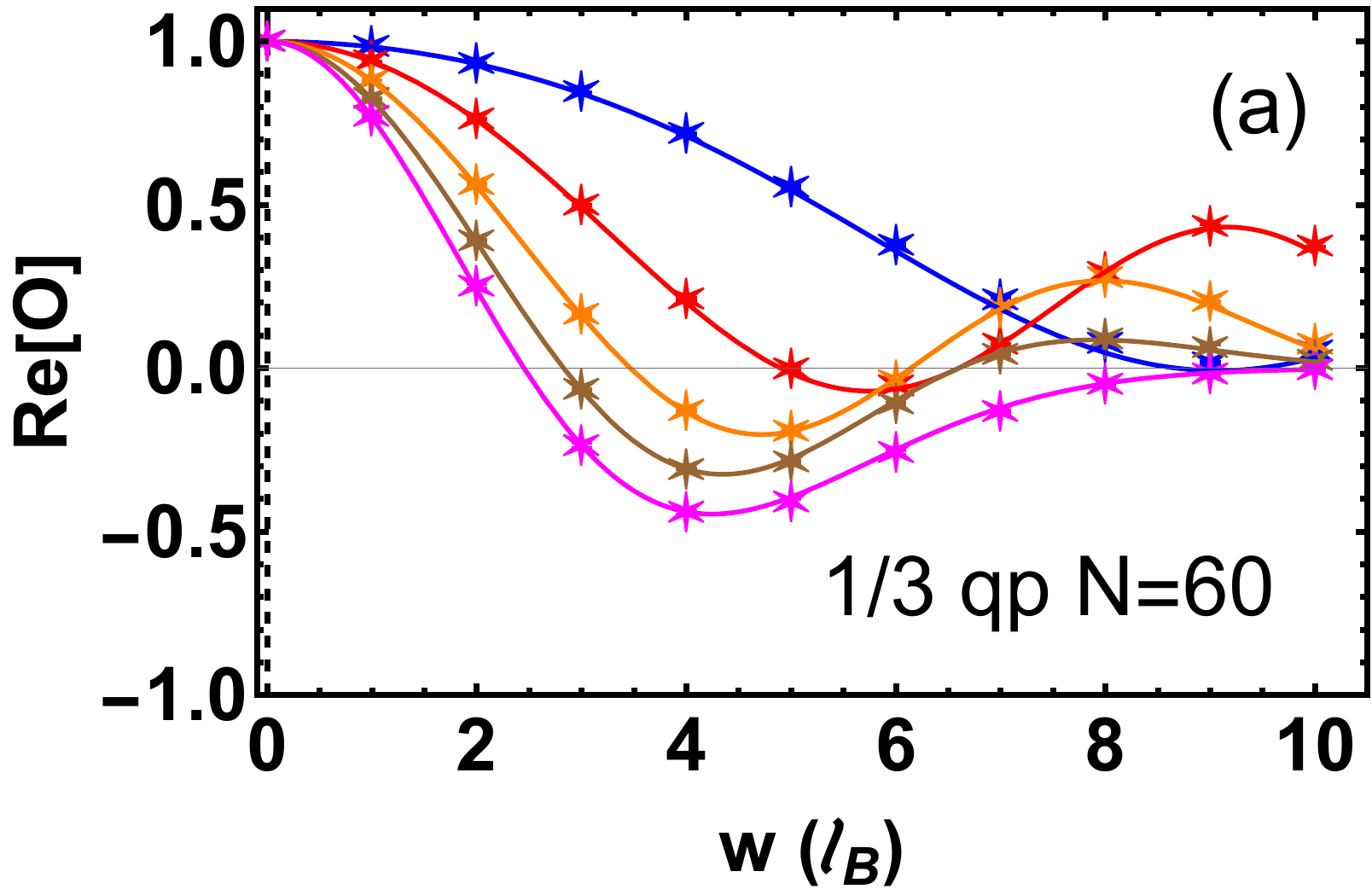} 
	\includegraphics[width=0.48\columnwidth]{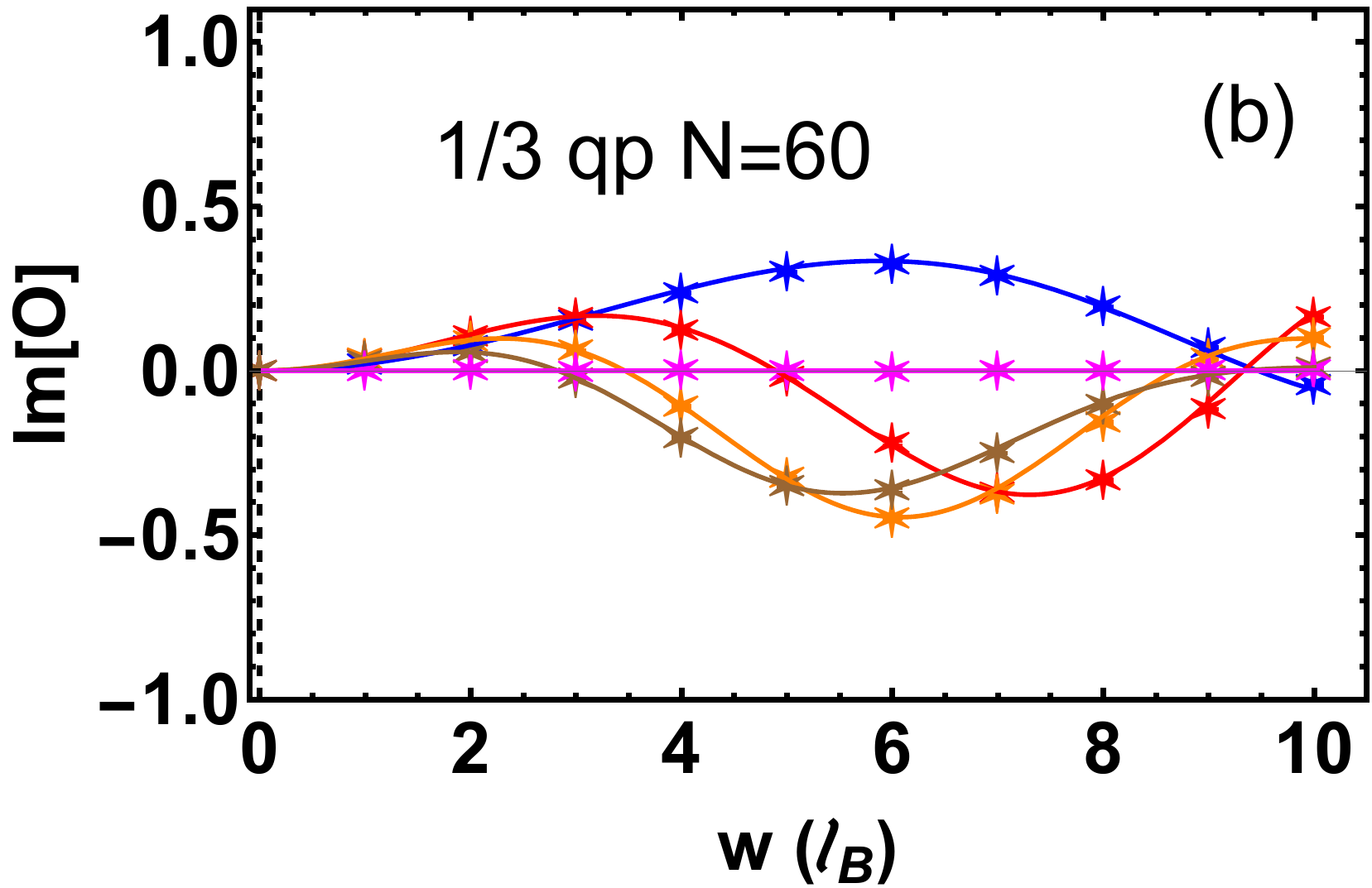} 
    \includegraphics[width=0.48\columnwidth]{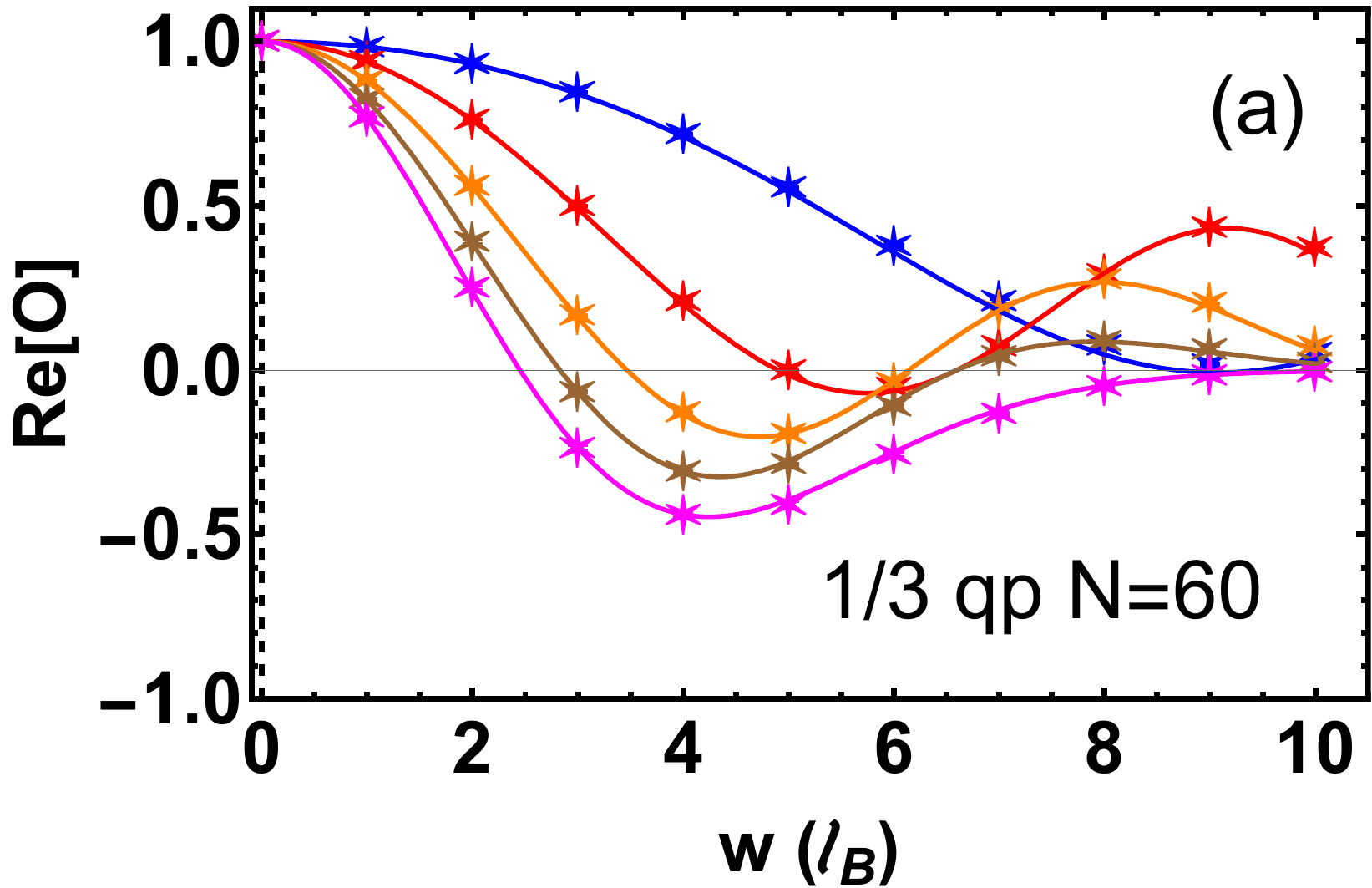} 
	\includegraphics[width=0.48\columnwidth]{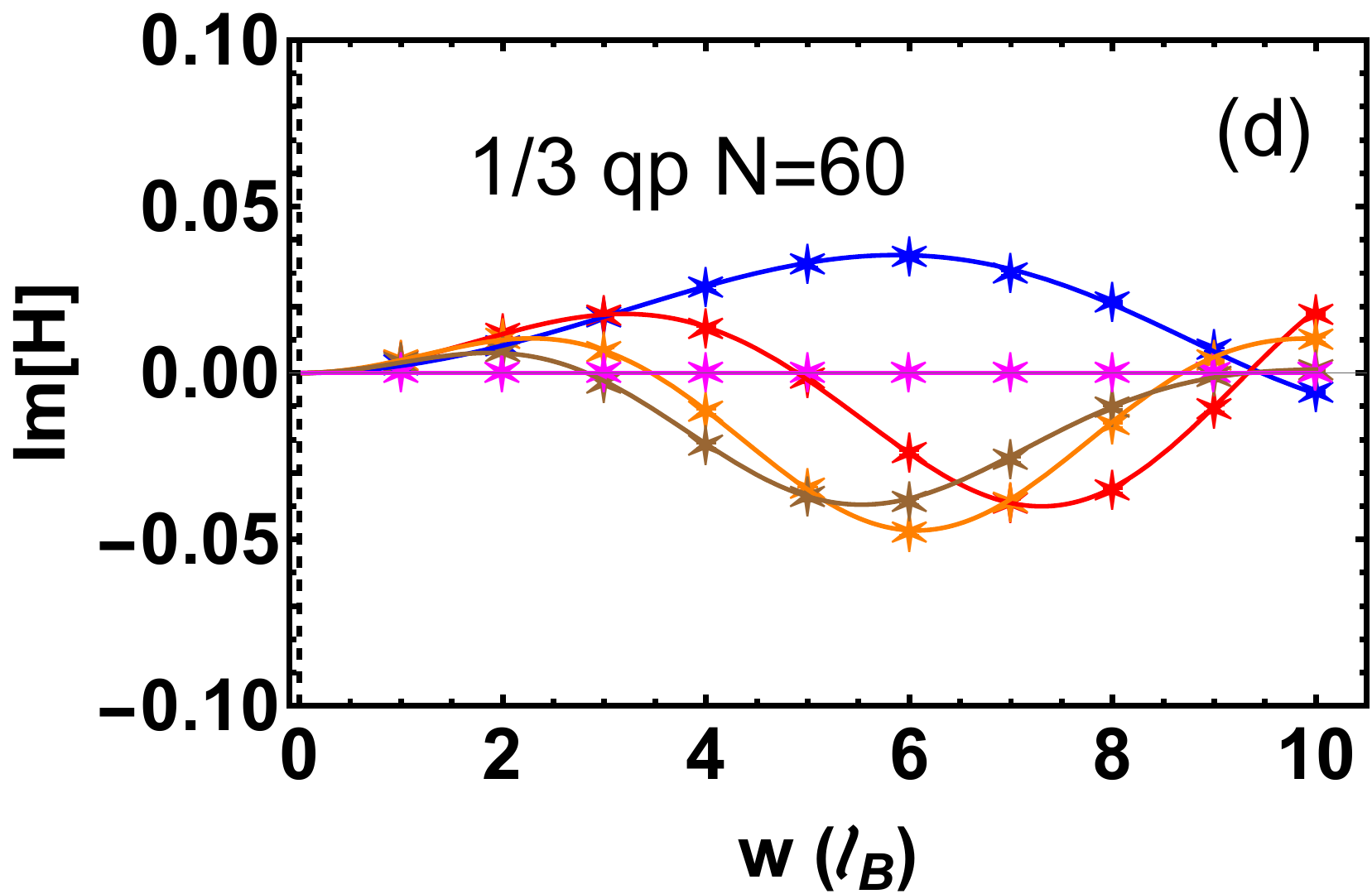} 
    \includegraphics[width=0.48\columnwidth]{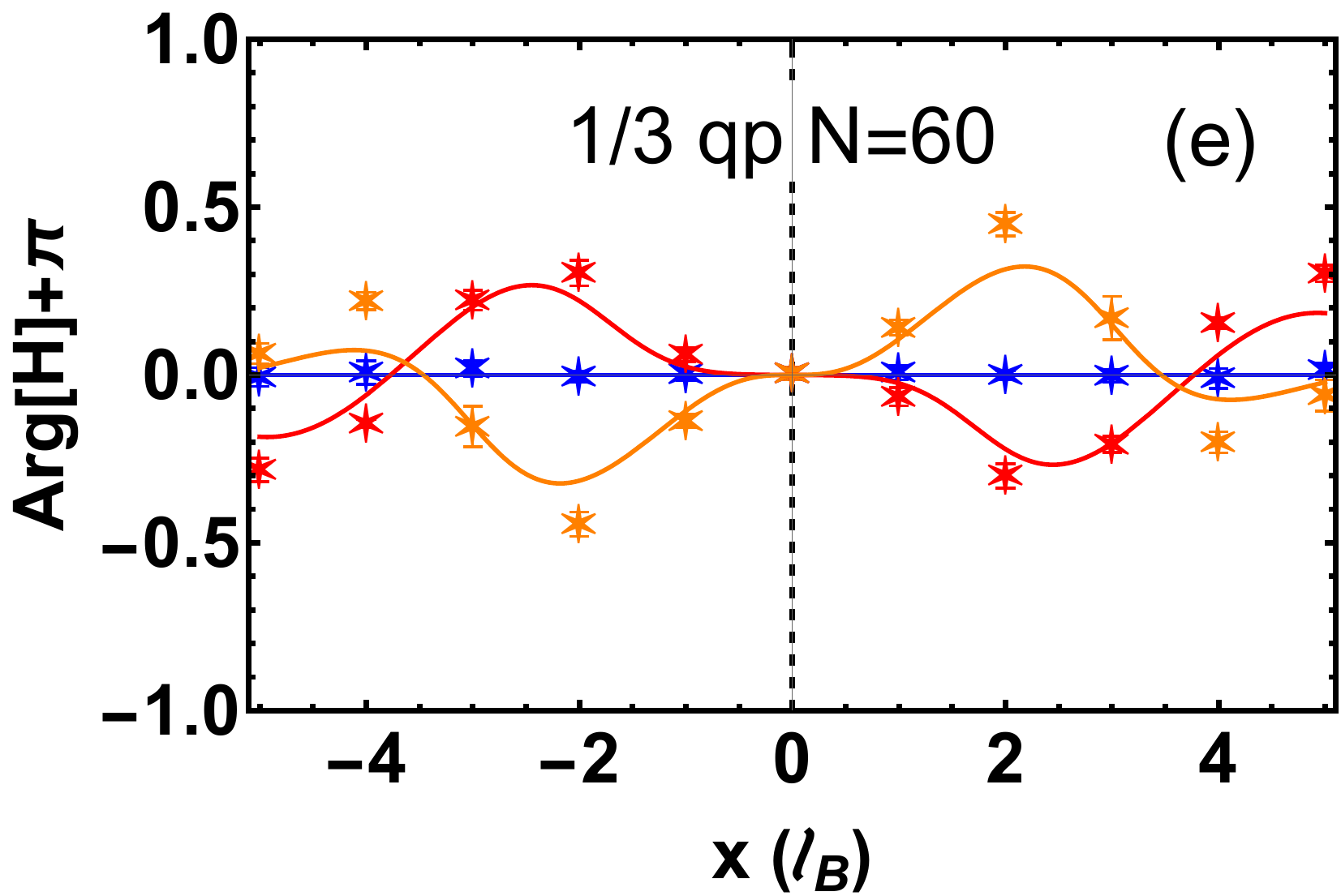} 
	\includegraphics[width=0.48\columnwidth]{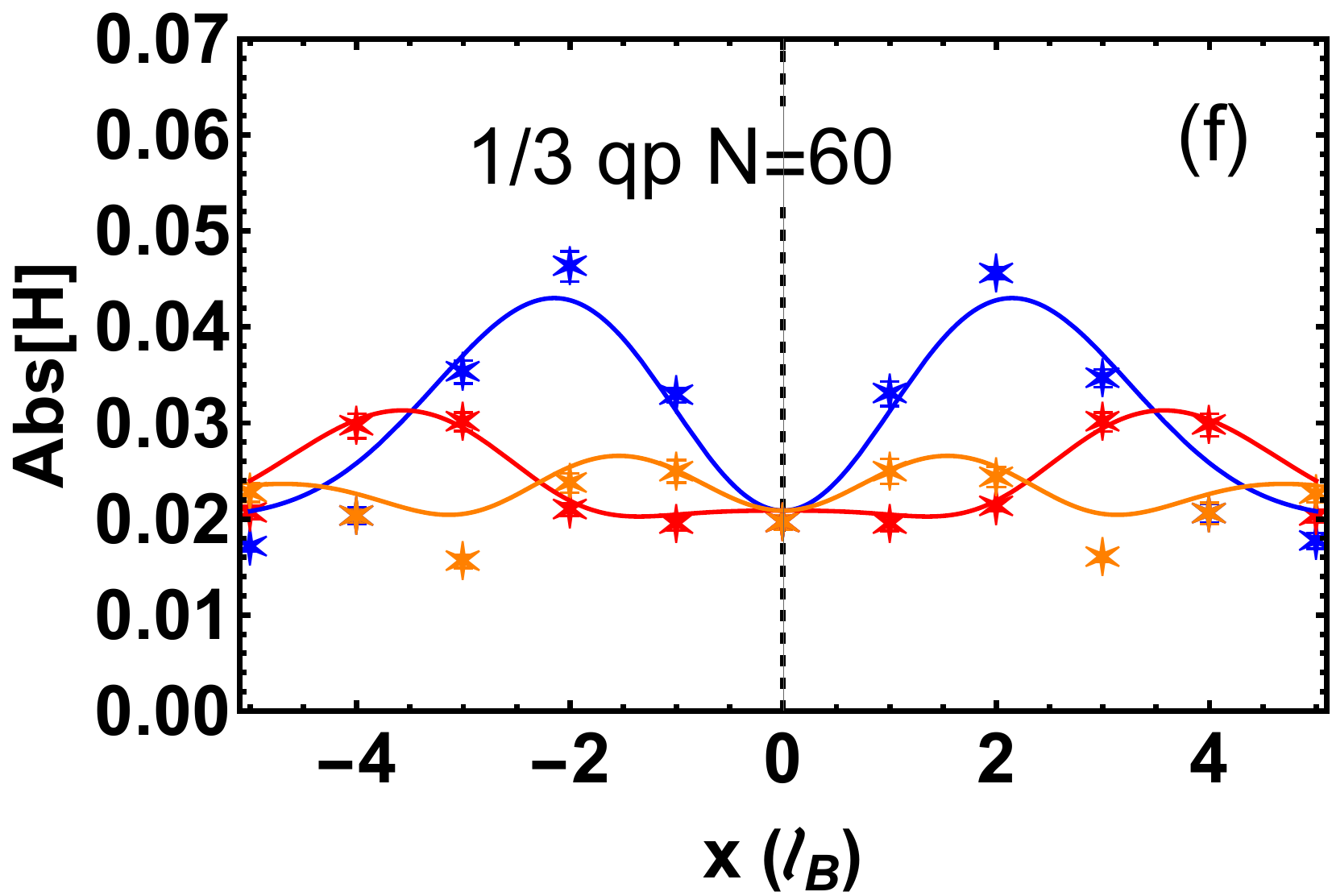} 
   \includegraphics[width=0.48\columnwidth]{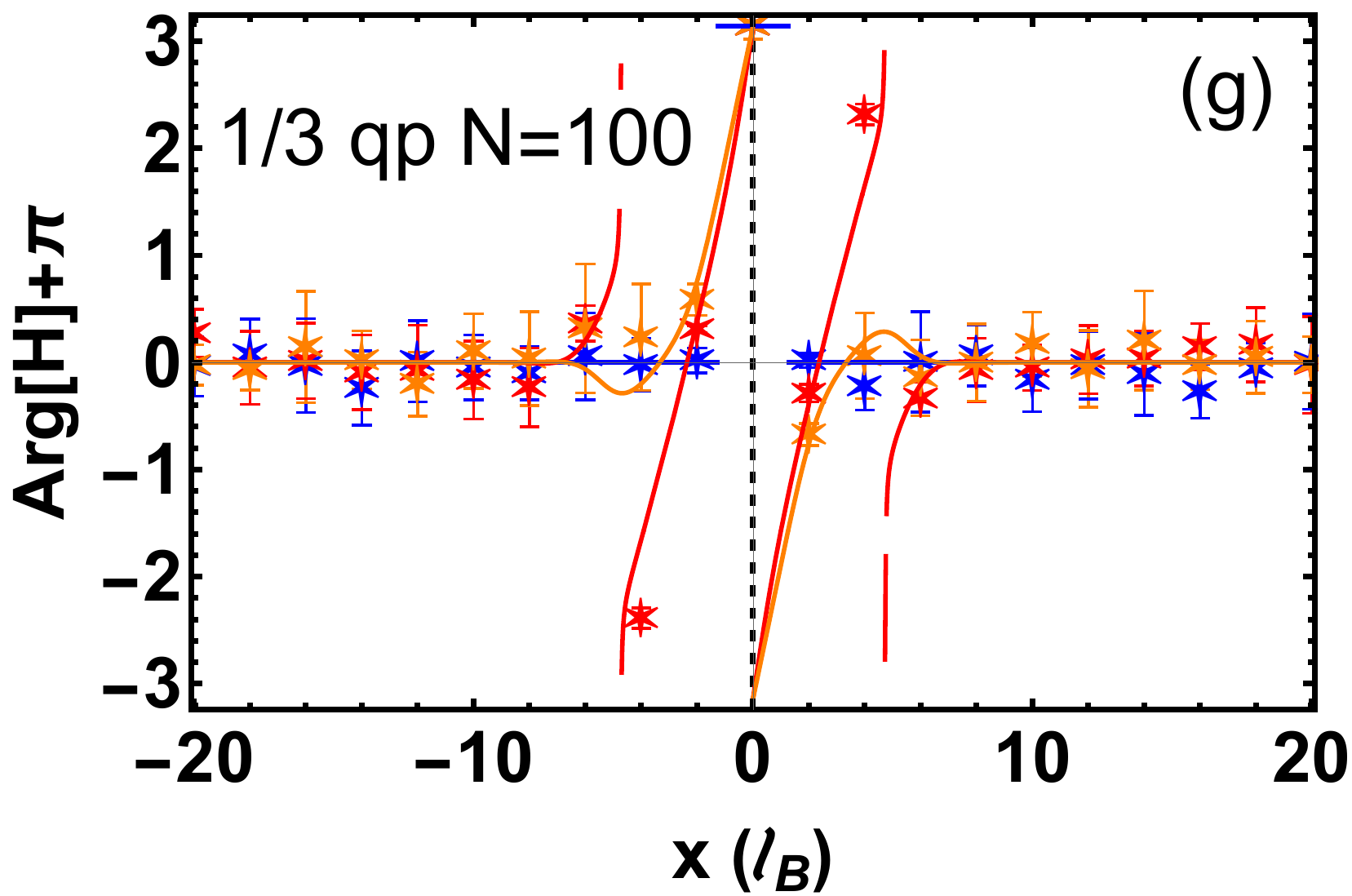} 
	\includegraphics[width=0.48\columnwidth]{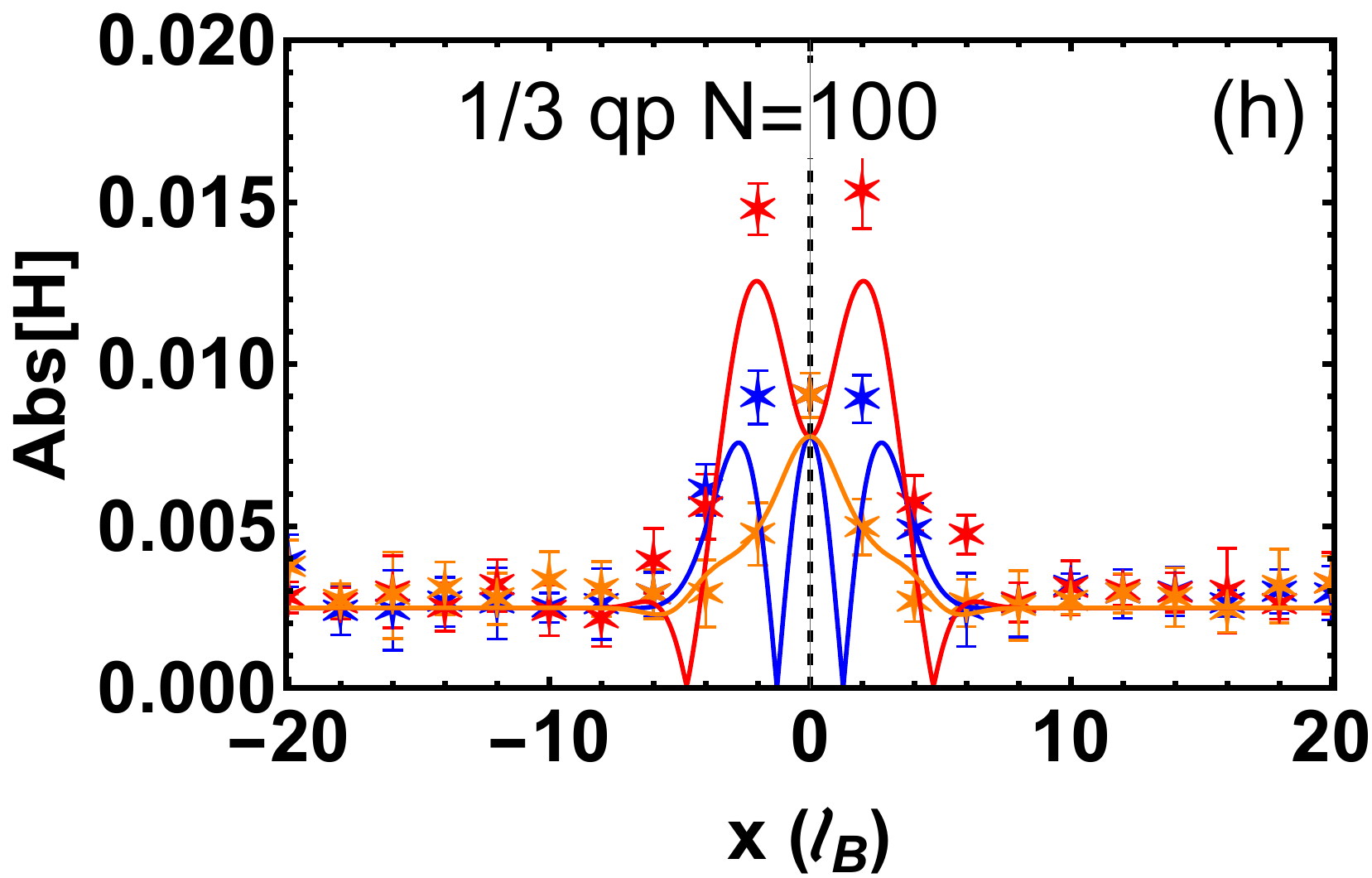} 
	\caption{Same as in Fig.~\ref{O25} but for the quasiparticle of $\nu=1/3$.}
	\label{O13}
\end{figure}

With the wave functions in hand, we first proceed to evaluate the overlap matrix $O^{{\rm qp}-n}_{ww'}=\langle \Psi_w^{{\rm qp}-n}|\Psi_{w'}^{{\rm qp}-n}\rangle$, assuming $2p=2$ below. 
The top two panels in Figs.~\ref{O25} and \ref{O13} display the real and imaginary parts of the matrix element $O^{{\rm qp}-2}_{ww'}$ for the quasiparticle of $\nu=2/5$ and $\nu=1/3$ (stars). For comparison, we also show the IQHE counterpart, 
$\IO^{\rm qp-n}_{ww'}$ 
namely the overlap an electron localized at $w$ and that with that at $w'$ (solid lines). This can be analytically evaluated to be (See Appendix \ref{overlap element})
\begin{equation}
\IO^{\rm qp-n}_{ww'} = \sqrt{2\pi}\;\eta_{0}^{(n)} (w-w',B^*)e^{\imath \frac{{\rm Im}(w\bar{w}')}{2\ell^{*2}}}.
\end{equation}
The correspondence between the overlaps for CF quasiparticle in the $n$th $\Lambda$L and for electron in the $n$th LL is strikingly close (the discussion in Appendix \ref{overlap element} provides further insight into this result).

We next come to the comparison of matrix elements of the Hamiltonian 
$H=\sum_{\alpha,k} \epsilon_k \delta(\vec{r}_\alpha-\vec{w}_k)$.  
The tunneling matrix element $H^{{\rm qp}-n}_{ij}=\langle\Psi^{{\rm qp}-n}_{w_i} | H
|\Psi^{{\rm qp}-n}_{w_j} \rangle$ is given by $H^{{\rm qp}-n}_{ij}=\epsilon_i H^{{\rm qp}-n}_{w_iw_j,w_i}+\epsilon_j H^{{\rm qp}-n}_{w_iw_j,w_j}+\sum_{k\neq i,j} \epsilon_k H^{{\rm qp}-n}_{w_iw_j,w_k}$, where $H^{{\rm qp}-n}_{w_i w_j,w_k}\equiv \langle\Psi_{w_i} | \sum_{\alpha}\delta(\vec{r}_\alpha-\vec{w}_k)|\Psi_{w_j}\rangle$. We consider separately the two types of terms: $H^{{\rm qp}-n}_{w_iw_j,w_i}$, and $H^{{\rm qp}-n}_{w_iw_j,w_k}$ with $k\neq i,j$. 
In panels (c) and (d) of Figs.~\ref{O25} and \ref{O13}, we show $H_{w_1w_2,w_1}$ for the quasiparticle at $\nu=2/5$ and $\nu=1/3$. Panels (e) - (h) show $H_{w_1 w_2,w}$ with $w\neq w_1,w_2$. In the numerical calculations, we have approximated the $\delta$ functions as normalized Gaussians of a small width $0.1\ell$. 
The corresponding tunneling matrix elements for the quasiparticle of the IQHE state can be calculated analytically as shown in the Appendix~\ref{tunneling} The explicit formula for $\IH^{{\rm qp}-n}_{w_1w_2,w}$ is:
\begin{multline}
\label{tunneling element}
\IH^{{\rm qp}-n}_{w_i w_j,w}={\nu\over \sqrt{2\pi}}\eta_0^{(n)}(w_i-w_j)e^{{i\over 2}{\rm Im}[\bar{w_j}w_i]}\\ 
+\eta_0^{(n)}(w-w_j)e^{{i\over 2}{\rm Im}[\bar{w_j}w]}\eta_0^{(n)}(w_i-w)e^{{i\over 2}{\rm Im}[\bar{w}w_i]}
\end{multline}
These are also shown in Figs.~\ref{O25} and \ref{O13} (solid lines). Again a close correspondence can be seen between the FQHE and the IQHE results. Appendix~\ref{additionalQPQH} also shows analogous results for the quasiholes at $\nu=1/3$ and $2/5$.

These studies provide a quantitative confirmation of the ``law of corresponding states"~\cite{Jain90b,Kivelson92}. We should remember that the wave functions in Eqs.~\ref{CFQP} and ~\ref{CFQH} are extremely accurate representations of the actual quasiparticles and quasiholes in the FQHE, and thus, the calculated $O_{ij}$ and $H_{ij}$ are accurate representations of the actual overlap and tunneling matrix elements for the exact quasiparticles and quasiholes. In other words, if an exact diagonalization study were possible for large system sizes, it would have yielded very nearly the same overlap and tunneling matrix elements.  While the CF theory provides a natural framework for understanding this correspondence, the quantitative closeness of the comparison was unexpected, given that in the FQHE we are dealing with the overlap and tunneling matrix element between two collective, many-particle states, and there is no a priori reason for these to correspond to overlap and tunneling matrix element of a single electron. The correspondence is particularly striking given that these quantities depend sensitively on the filling factor as well as on the distance across which a quasiparticle or a quasihole tunnels.

{\it Implications for experiments:}  
The above correspondence between the overlap matrix elements implies that the problem of Anderson localization of a quasiparticle of the $\nu=n/(2pn+1)$ FQHE state is essentially identical to that of the Anderson localization of an electron in the $\nu=n$ IQHE state,  provided that the onsite binding energies are appropriately rescaled. These two systems are therefore predicted to have identical critical properties. We can thus export {\it mutatis mutandis} the results from the study of non-interacting electrons in a magnetic field to make predictions for the nature of localization in the FQHE. We list these, along with the underlying assumptions.

For a single electron in a Landau level, it is believed that the localization length diverges at some critical energy $E_c$. Furthermore, the divergence of the localization length as $E_c$ is approached is described, at least approximately, by $\xi\sim |E-E_c|^{-\alpha}$ where $\alpha$ is the effective localization length exponent. These features have found support from direct numerical studies that solve the Schr\"odinger equation on a disordered strip of finite width, and then use finite size scaling to deduce the localization length in the thermodynamic limit. 
The correspondence established above implies that the localization length of a single CF quasiparticle also diverges at a critical energy, with the same exponent as that for electrons. 
Numerical calculations have shown that the divergence of the localization length within the lowest Landau level (relevant to $1\rightarrow 2$ transition) is characterized by an exponent $\alpha\approx 2.3-2.5$ \cite{Huckestein92,Huo92}.  In the second LL, where the numerical calculations are less reliable, a higher value $\alpha\approx 5.5$ is obtained for short range disorder~\cite{Mieck93,Liu93,Liu94}, and the exponent has been found to depend on the range of the disorder.   It has been argued~\cite{Huckestein94} that in higher LLs there is a large irrelevant length and only for system sizes and localization lengths larger than this irrelevant length, not readily accessible to numerics, can a single parameter scaling behavior be observed. 
It is generally believed that in the asymptotic, critical regime, the localization length exponent is $\alpha\approx 2.3-2.5$ in all LLs. This is consistent with calculations using a quantum percolation model that allows tunneling across saddle points~\cite{Chalker88,Lee93}. This value is also consistent with that measured in the most reliable experiments~\cite{Li09}.
Translating these results to composite fermions, we predict $\alpha\approx 2.3-2.5$ also for the FQHE.

So far we have considered a single quasiparticle. Increasing the electron filling factor amounts to creating a finite density of CF quasiparticles in the topmost $\Lambda$L. The problem now is more complex because one must now deal with the interaction between composite fermions as well as  their fractional braiding statistics. We assume that composite fermions are weakly interacting. This may be justified given that a large portion of the Coulomb interaction has been spent into creating composite fermions, which are themselves much more weakly interacting than electrons; the interaction pseudopotentials for composite fermions are reduced by an order of magnitude or more relative to the Coulomb pseudopotentials for electrons~\cite{Lee01,Lee02}. Alternatively, one can note that the charge associated with the CF quasiparticle or CF quasihole has magnitude $e/(2pn\pm 1)$, leading to an inter-CF interaction that is reduced by a factor of $(2pn\pm 1)^{-2}$ relative to the Coulomb interaction between electrons. The fractional braiding statistics can be incorporated by mapping into an IQHE problem with the effective magnetic field given by $B^*=B-2\pi\rho(\vec{r})$ where the spatial dependence of the density $\rho(\vec{r})$ encodes information on fractional braiding statistics. One must find eigenstates of ``non-interacting" composite fermions in the presence of disorder, which must be done in a self-consistent manner as the Hamiltonian depends on the density~\cite{Hu19} (composite fermions are not really non-interacting, as the solution for any eigenstates depends on other eigenstates). The problem is simplified if we make the mean field approximation of replacing $\rho(\vec{r})$ by its average value, and thus treat composite fermions in a uniform $B^*$. It appears to us, although we cannot prove it, that this approximation is likely to become valid for single-CF wave functions whose extent is much larger than the separation between quasiparticles; such wave functions with large localization lengths are of primary interest in the vicinity of the phase transition. (As shown in Refs.~\cite{Canright89b,Pryor91}, a charged particle in the presence of a lattice of fractional flux tubes behaves similarly to a particle in a uniform magnetic field.) Then the problem maps into that of weakly interacting electrons in a uniform effective magnetic field, and one can translate the above results into a behavior as a function of the filling factor, because increasing the filling factor simply fills successively higher energy single particle CF orbitals. This implies the power law behavior $\xi\sim |\nu-\nu_c|^{-\alpha}$. 

Within the mean field approximation, for symmetric disorder, the extended state is predicted to occur at the critical filling factor $\nu_c^*=n+1/2$, which determines the position of the peak separating the incompressible states at $\nu^*=n$ and $\nu^*=n+1$ to be $\nu_c=(n+1/2)/[2p(n+1/2)\pm 1]$. There is experimental support for this prediction~\cite{Goldman90b}.

Field theoretical treatments have suggested universality of the localization physics in the FQHE and the IQHE~\cite{Laughlin85,Kivelson92b,Burgess01,Huckestein96}. Our work provides microscopic support for it.

In summary, we have shown that in spite of the strongly correlated nature of the FQHE, it is possible to make detailed quantitative predictions regarding the nature of localization, exploiting the observation that 
the analogy between the FQHE and the IQHE also extends to the localization physics.

The work at Penn State (S.P. and J.K.J) was supported in part by the U. S. Department of Energy, Office of Basic Energy Sciences, under Grant No. DE-SC0005042. The numerical part of this research was conducted with Advanced CyberInfrastructure computational resources provided by the Institute for CyberScience at the Pennsylvania State University. GJS acknowledges DST/SERB grant ECR/2018/001781 and National Supercomputing Mission (NSM) for providing computing resources of PARAM Brahma at IISER Pune, which is implemented by C-DAC and supported by the Ministry of Electronics and Information Technology (MeitY) and Department of Science and Technology (DST), Government of India.  

\begin{appendix}

\section{Overlaps between quasiparticle / quasihole states and momentum independence of norms}
\label{overlap element}
The surprising similarity between the overlaps $O_{w_1w_2}$ for the quasiparticle / quasihole of a FQHE state with the corresponding overlaps $\mathcal{O}_{w_1w_2}$ for the particle / hole of an IQHE state 
can be traced back to the momentum independence of the norm of the single quasiparticle / quasihole states.

First consider a single QH located at $w$ in an IQH state at a field $B^*$. For concreteness of the discussion, we shall describe the QH of the $\nu=2$ case here. The wave function in the disc geometry can be written as 
\begin{multline}
\psi^{{\rm qh}-2}_w[z_i,B^*]=\\\det \left[\begin{array}{cccc}
0 & \eta^{(0)}_{0}\left(z_{1}\right) & \eta^{(0)}_{0}\left(z_{2}\right) & \dots\\
0 & \eta^{(0)}_{1}\left(z_{1}\right) & \eta^{(0)}_{1}\left(z_{2}\right) & \dots\\
\vdots & \vdots & \vdots & \ddots\\
\eta^{(1)}_{-1}\left(w\right) & \eta^{(1)}_{-1}\left(z_{1}\right) & \eta^{(1)}_{-1}\left(z_{2}\right) & \dots\\
\eta^{(1)}_{0}\left(w\right) & \eta^{(1)}_{0}\left(z_{1}\right) & \eta^{(1)}_{0}\left(z_{2}\right) & \dots\\
\vdots & \vdots & \vdots & \ddots\\
\eta^{(1)}_{N_{1}-2}\left(w\right)& \eta^{(1)}_{N_{1}-2}\left(z_{1}\right) & \eta^{(1)}_{N_{1}-2}\left(z_{2}\right) & \dots
\end{array}\right]\nonumber
\end{multline}

\begin{figure}[!htb]
\includegraphics[width=\columnwidth]{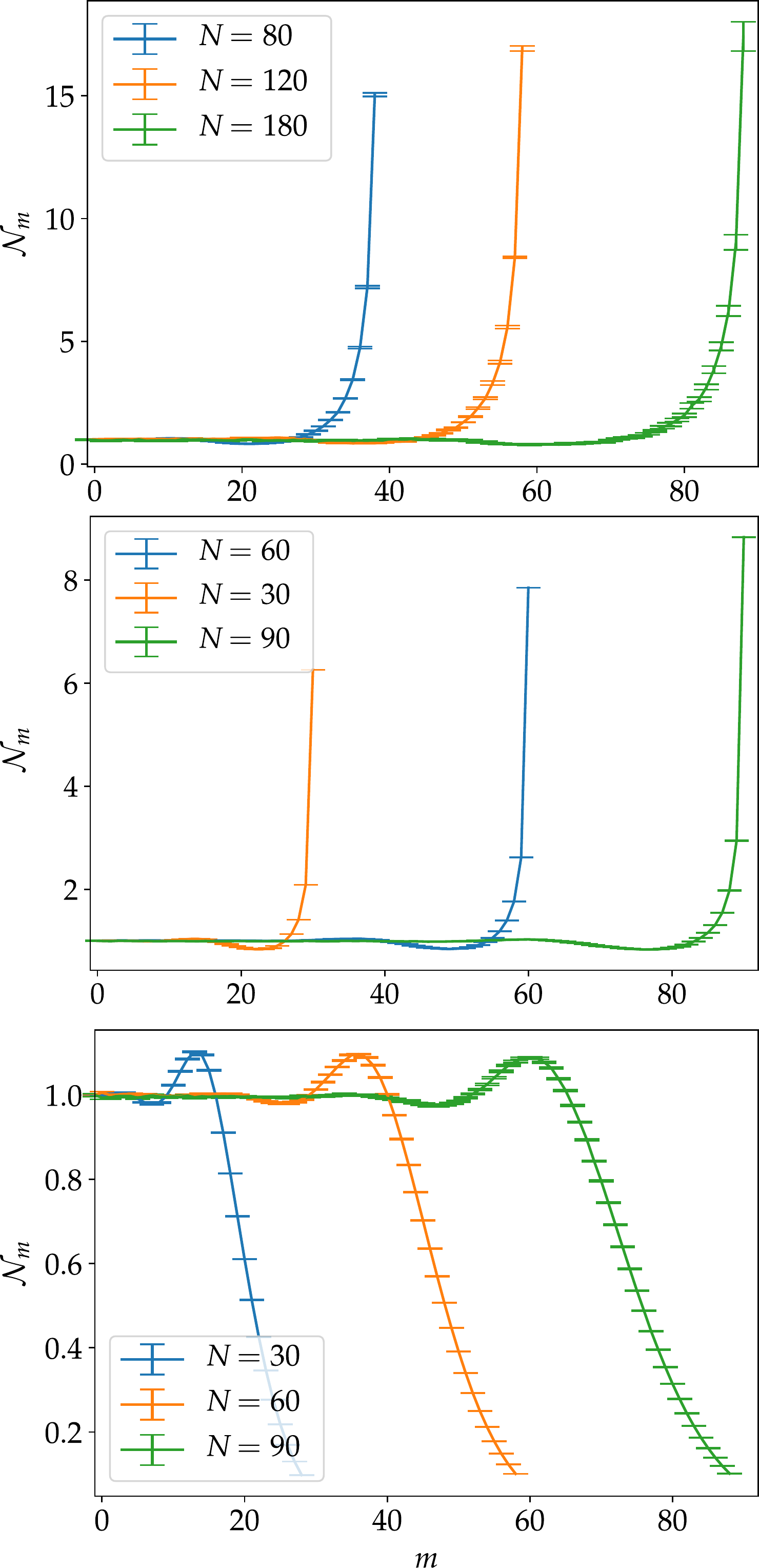}
\caption{Explicit evaluation of the norm $\mathcal{N}_m$ defined in Eq.~\ref{eq:norm}.
The norm is independent of the momentum $m$ for a range of momenta $m$ which contribute to the coherent states of these excitations in the bulk. The three panels show the norms of single quasihole states of $2/5$ (top), single quasihole state of $1/3$ (middle) and single quasiparticle state of $1/3$ (bottom). Different lines indicate different electron numbers. 
\label{fig:norms}}
\end{figure}

Here $\eta^{(n)}_{m}(z)\equiv\eta^{(n)}_{m}(z,B^*)$ is the normalized single particle wave function of an electron in the angular momentum $m$ orbital in the $n^{\rm th}$ Landau level (with $n=0$ labeling the LLL) and $N_1$ is the number of electrons in the second Landau level. By expanding the determinant along the first column, we can re-write the wave function as follows
\begin{equation}
\psi^{{\rm qh}-2}_w[z_i,B^*]=\sum_{m=-1}^{N_1-2} (-1)^m \eta^{(1)}_{m}(w,B^*) \phi^{(2)}_m(z,B^*) 
\end{equation}
where $\phi^{(2)}_m(z)$ is the wave function of two filled LLs except for a hole in the orbital with angular momentum $m$.
We can write the overlap of two such states $\left | w_1 \right \rangle\equiv\psi^{{\rm qh}-2}_{w_1}$ and $\left | w_2 \right \rangle\equiv\psi^{{\rm qh}-2}_{w_2}$ as
\begin{gather}
\IO^{{\rm qh}-2}_{w_1w_2}=\frac{\left \langle w_1 | w_2\right \rangle}{\sqrt{\left \langle w_1 | w_1\right \rangle \left \langle w_2 | w_2\right \rangle}}\label{eq:ow1w2}\\
\left \langle w_1 | w_2\right \rangle=\sum_{m=-1}^{N_1-2} \overline{\eta}_{1,m}(w_1,B^*)\eta_{1,m}(w_2,B^*)  \langle \phi^{(2)}_m | \phi^{(2)}_m  \rangle \nonumber
\end{gather}
where $ \langle \phi^{(2)}_m | \phi^{(2)}_m  \rangle$ is the norm of the momentum state. The norm is $N!$, which, crucially, is independent of $m$. For a sufficiently large number of particles, and $w_i$ deep inside the bulk (away from the boundary),
we may send the upper limit of the summation shown above from $N_1-2$ to infinity; which allows us to explicitly evaluate the sum and get $\IO^{{\rm qh}-2}_{w_1w_2}=\sqrt{2\pi}e^{\frac{1}{2} {\rm Im}(\overline{w_1}w_2)} \eta^{(1)}_{0}(w_1-w_2)$. It can be shown in a similar manner that the overlap in the case of qh at filling fraction $n$ is 
\begin{equation}
\IO^{{\rm qh}-n}_{w_1w_2} =  \sqrt{2\pi}e^{\frac{{\rm Im}(\overline{w}_1w_2)}{2\ell^{*2}} } \eta^{(n-1)}_{0}(w_1-w_2,B^*).
\end{equation}

Next we consider the corresponding calculation at filling fraction $2/5$ which occurs at $B=5B^*$.  
The state $\psi^{{\rm qh}-2/5}_w[z_i]$ of a single quasihole located at $w$ is obtained by composite fermionising the IQH wave function $\psi^{{\rm qh}-2}_w[z_i]$ at the effective magnetic field. This wave function is given by 
\begin{equation}
\psi^{{\rm qh}-2/5}_w[z_i,B]=\mathcal{P}_{\rm LLL} J^2 \psi^{{\rm qh}-2}_{w}[z_i,B^*]
\end{equation}
where $J^2=\left[\prod_{i<k=1}^N(z_i-z_k) e^{-\sum_j|z_j|^2/4\ell_1^2}\right]^{2}$.

Following a similar sequence of steps as in the IQHE case, we arrive at a result analogous to Eqs \ref{eq:ow1w2}
\begin{gather}
O^{{\rm qh}-2/5}_{w_1w_2}=\frac{\left \langle w_1 | w_2\right \rangle}{\sqrt{\left \langle w_1 | w_1\right \rangle \left \langle w_2 | w_2\right \rangle}}\label{eq:o25w1w2}\\
\left \langle w_1 | w_2\right \rangle=\sum_{m=-1}^{N_1-2} \overline{\eta}_{1,m}(w_1,B^*)\eta_{1,m}(w_2,B^*)  \langle \phi^{2/5}_m | \phi^{2/5}_m  \rangle \nonumber
\end{gather}
where $\phi^{2/5}_m(z,B)=\mathcal{P}_{\rm LLL} J^2 \phi^{(2)}_m(z,B^*)$.

Here, we make the observation that if the norm 
\be
\mathcal{N}_m=  \langle \phi^{2/5}_m | \phi^{2/5}_m  \rangle 
\label{eq:norm}
\ee
 is independent of the momentum, results similar to the IQH case follow. 
Indeed, as shown in Fig \ref{fig:norms} we find, from explicit evaluation, that for momenta that contributes to the localized QH states well inside the bulk of the FQH state, the $\mathcal{N}_m$ is independent of $m$. 
For quasiholes located deep in the bulk, we may then write the overlap as 
\begin{equation}
O^{{\rm qh}-2/5}_{w_1w_2}(B) =  \sqrt{2\pi}e^{\frac{{\rm Im}(\overline{w}_1w_2)}{2\ell^{*2}} } \eta^{(1)}_{0}(w_1-w_2,B^*).
\end{equation}
The relation $\IO^{{\rm qh}-2/5}_{w_1w_2}=O^{{\rm qh}-2}_{w_1w_2}$ 
is thus related to the momentum independence of the norms $\mathcal{N}_m$ of momentum eigenstates of quasiholes. Though the above discussion is restricted to a QH, an analogous result is found to hold also for the quasiparticle state of $1/3$.

\section{Tunneling matrix elements of the IQHE}
\label{tunneling}
In this section, we analytically derive the tunneling matrix elements of the IQHE, {\it i.e.}, Eq.~\ref{tunneling element} in the main text. We first show this for quasiparticle states with $n$ filled Landau levels and a quasiparticle localized in the $n$th Landau level. The tunneling matrix element is
\be
\label{tunneling element1}
\IH_{w_i w_j,w}^{{\rm qp}-n}=\langle\Psi^{{\rm qp}-n}_{w_i} |\hat{\rho}(w)|\Psi^{{\rm qp}-n}_{w_j} \rangle
\ee
where the density operator $\hat{\rho}(w)=\sum_i\delta(\vec{r}_i-w)=2\pi\sum_{n_1,n_2,m_1,m_2}\bar{\eta}_{m_1}^{(n_1)}(w)\eta_{m_2}^{(n_2)}(w)c^{(n_1)\dagger}_{m_1}c^{(n_2)}_{m_2}$ and $|\Psi^{{\rm qp}-n}_{w_j} \rangle=\sum_{m_3}\bar{\eta}_{m_3}^{(n)}(w_j)c^{(n)\dagger}_{m_3}|\Psi_g^{(n)}\rangle$ in the second quantization notations. ($|\Psi_g^{(n)}\rangle$ is the ground state with $n$ LLs filled.) It follows that
\be
\IH_{w_i w_j,w}^{{\rm qp}-n}=2\pi\sum_{m_3,m_4}\eta_{m_4}^{(n)}(w_i)\bar{\eta}_{m_3}^{(n)}(w_j)\rho_{m_3m_4}^{{\rm qp}-n}(w)
\ee
where 
\begin{widetext}
\ba
\rho_{m_3m_4}^{{\rm qp}-n}(w)&=&\sum_{n_1,n_2,m_1,m_2}\bar{\eta}_{m_1}^{(n_1)}(w){\eta}_{m_2}^{(n_2)}(w)\langle\Psi_g^{(n)}|c_{m_4}^{(n)}c_{m_1}^{(n_1)\dagger}c_{m_2}^{(n_2)}c_{m_3}^{(n)\dagger}|\Psi_g^{(n)}\rangle\\ \nonumber
&=&
\begin{cases}
{\nu\over 2\pi}+|\eta_{m_3}^{(n)}(w)|^2 \quad\mbox{if } m_3=m_4\\
\bar{\eta}_{m_4}^{(n)}(w){\eta}_{m_3}^{(n)}(w) \quad\mbox{if } m_3\neq m_4
\end{cases}
\ea
where we assume that $w$ is deep in the bulk region such that $\sum_{n_1=0}^{n-1}\sum_{m_1=-n_1}^{\infty}|{\eta}_{m_1}^{(n_1)}(w)|^2={\nu\over 2\pi}$.

With this result and the identity
\be
\sum_{m=-n}^{\infty}\bar{\eta}_m^{(n)}(w_1){\eta}_m^{(n)}(w_2)={1\over\sqrt{2\pi}}\eta_0^{(n)}(w_2-w_1)e^{{i\over 2}{\rm Im}[\bar{w_1}w_2]}
\ee
we finally obtain
\ba
\IH_{w_i w_j,w}^{{\rm qp}-n}&=&\nu\sum_{m_3}\eta_{m_3}^{(n)}(w_i)\bar{\eta}_{m_3}^{(n)}(w_j)+2\pi\sum_{m_3}\eta_{m_3}^{(n)}(w)\bar{\eta}_{m_3}^{(n)}(w_j)\eta_{m_4}^{(n)}(w_i)\bar{\eta}_{m_4}^{(n)}(w)\\ \nonumber
&=&{\nu\over \sqrt{2\pi}}\eta_0^{(n)}(w_i-w_j)e^{{i\over 2}{\rm Im}[\bar{w_j}w_i]}+\eta_0^{(n)}(w-w_j)e^{{i\over 2}{\rm Im}[\bar{w_j}w]}\eta_0^{(n)}(w_i-w)e^{{i\over 2}{\rm Im}[\bar{w}w_i]}
\ea
which is Eq.~\ref{tunneling element} in the main text. Similarly, we have the following equation for the quasihole of $n$ filled LLs:
\ba
\IH_{w_i w_j,w}^{{\rm qh}-n}
&=&{\nu\over \sqrt{2\pi}}\eta_0^{(n-1)}(w_j-w_i)e^{{i\over 2}{\rm Im}[\bar{w}_iw_j]}-\eta_0^{(n-1)}(w-w_i)e^{{i\over 2}{\rm Im}[\bar{w}_iw]}\eta_0^{(n-1)}(w_j-w)e^{{i\over 2}{\rm Im}[\bar{w}w_j]}
\ea.
 \end{widetext}
 
\section{Results for quasiholes}
\label{additionalQPQH}
The wave function for a single CF quasihole in the second $\Lambda$L localized at $w$ for $\nu=2/(4p+1)$ is given by
\begin{eqnarray}
\Psi^{\rm qh-2}_{w} &=& {\cal P}_{\rm LLL} 
\left|\begin{array}{ccc}
\eta^{(1)}_{-1}(w,B^*) & \eta^{(2)}_{-1}(z_1,B^*) &\ldots\\
\eta^{(1)}_{0}(w,B^*) & \eta^{(2)}_{0}(z_1,B^*) &\ldots\\
\vdots&\vdots&\ldots \\
\eta^{(1)}_{N_1-2}(w,B^*) & \eta^{(2)}_{N_1-2}(z_1,B^*) &\ldots\\
0 & \eta^{(0)}_{0}(z_1,B^*) &\ldots\\
0 & \eta^{(0)}_{1}(z_1,B^*) &\ldots\\
\vdots&\vdots&\ldots \\
0 & \eta^{(0)}_{N_0-1}(z_1,B^*) &\ldots\\
\end{array}
\right| \nonumber \\
& & \;\;\;\;\times  \left[\prod_{i<k=1}^N(z_i-z_k) e^{-\sum_j|z_j|^2/4\ell_1^2}\right]^{2p}\;.
\label{CFQH}
\end{eqnarray}
Here the Slater determinant contains a single hole in the second LL localized at $z$.  The total number of particles is $N=N_0+N_1-1$.  Wave functions for quasiholes at other fractions can be constructed analogously. 

\begin{figure}[H]
	\includegraphics[width=0.48\columnwidth]{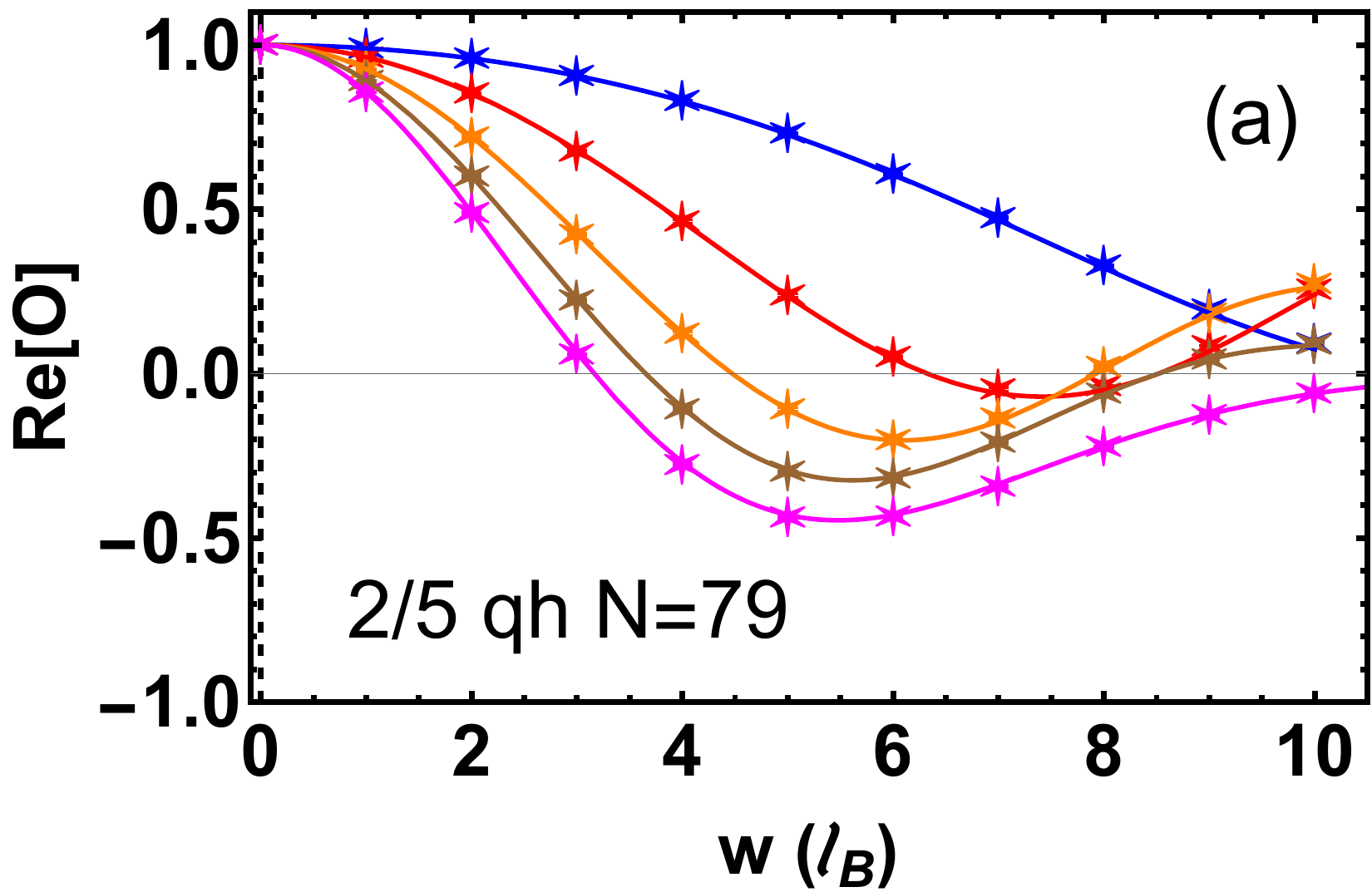} 
	\includegraphics[width=0.48\columnwidth]{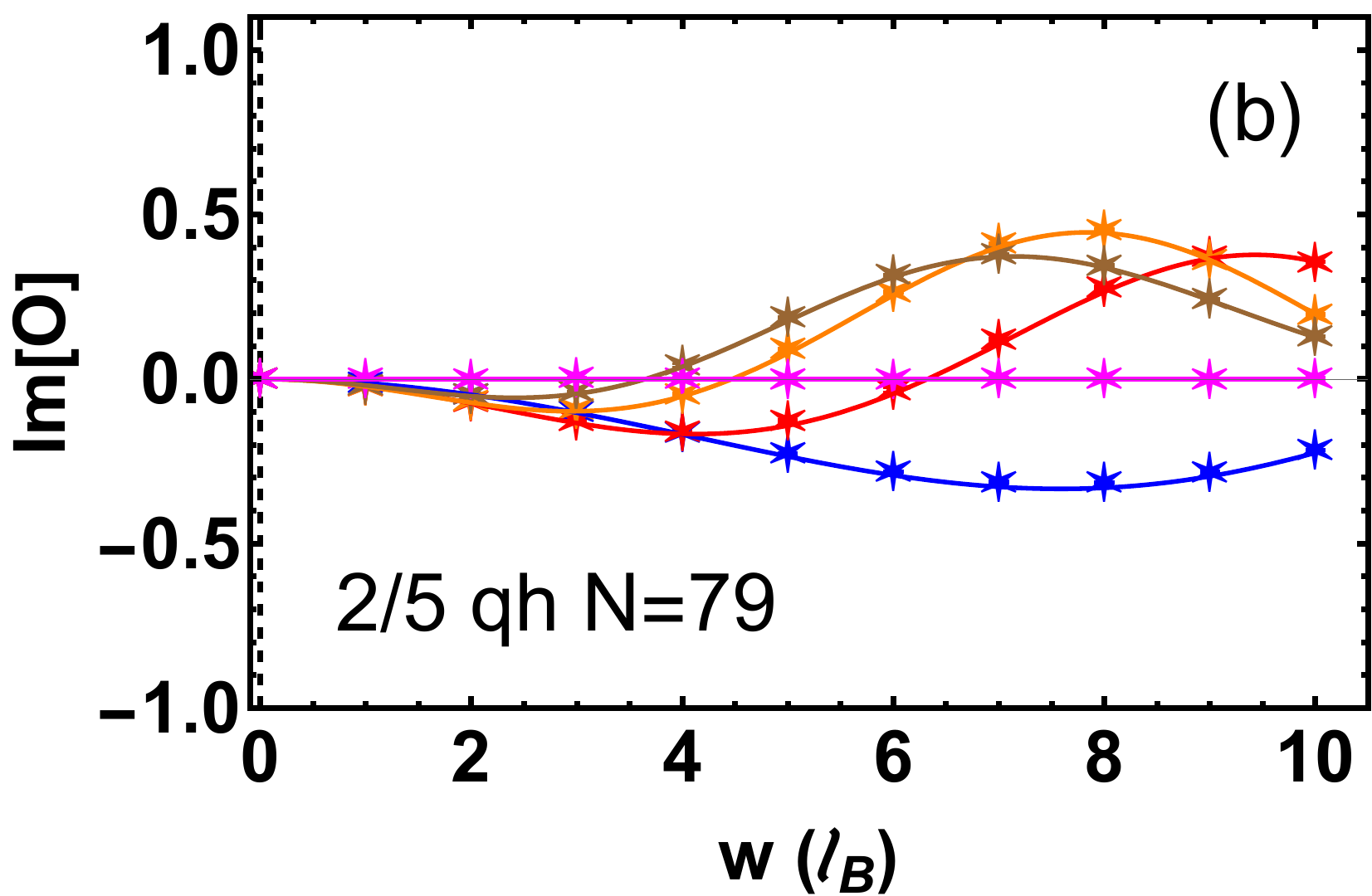} 
    \includegraphics[width=0.48\columnwidth]{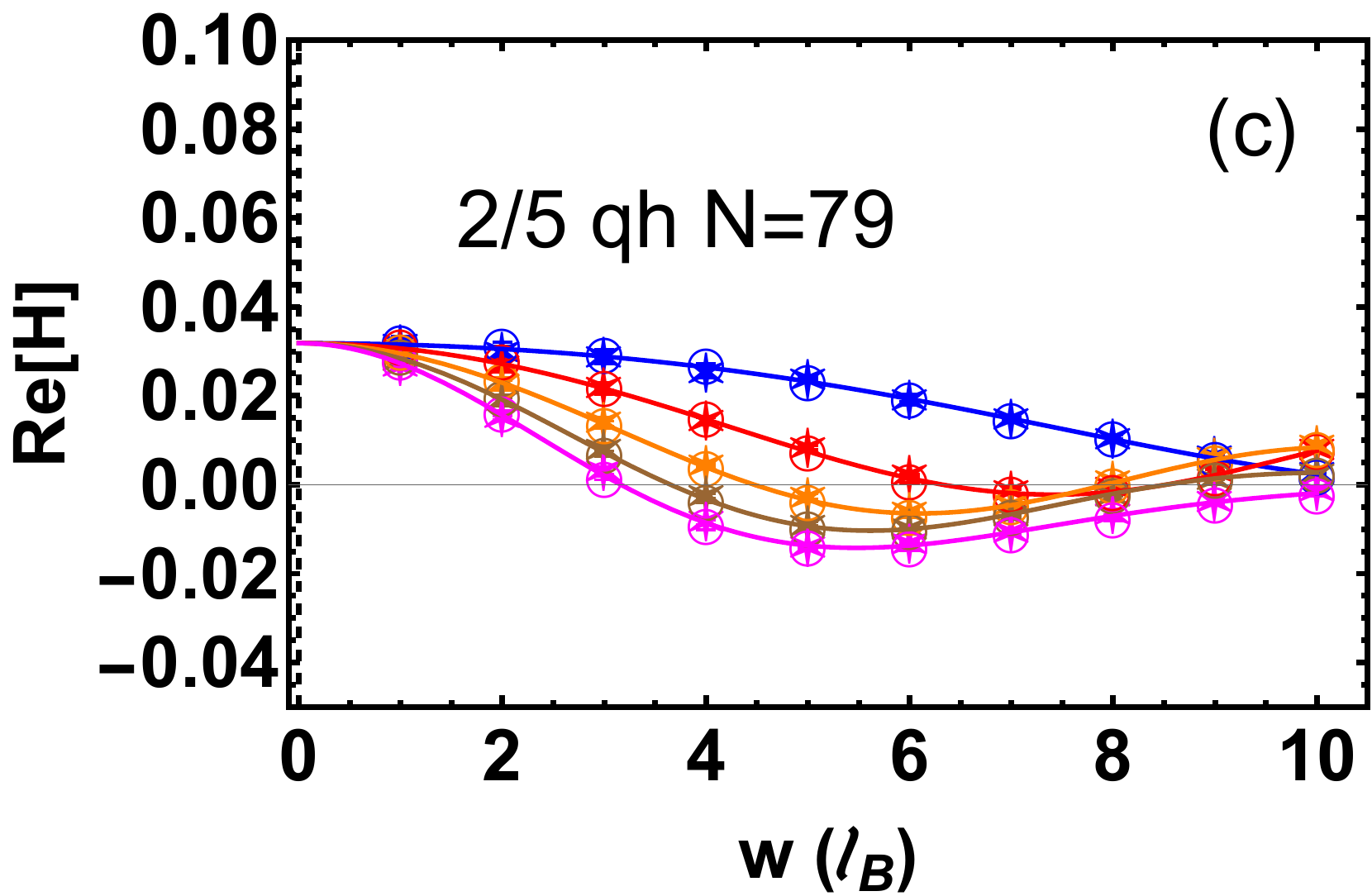} 
	\includegraphics[width=0.48\columnwidth]{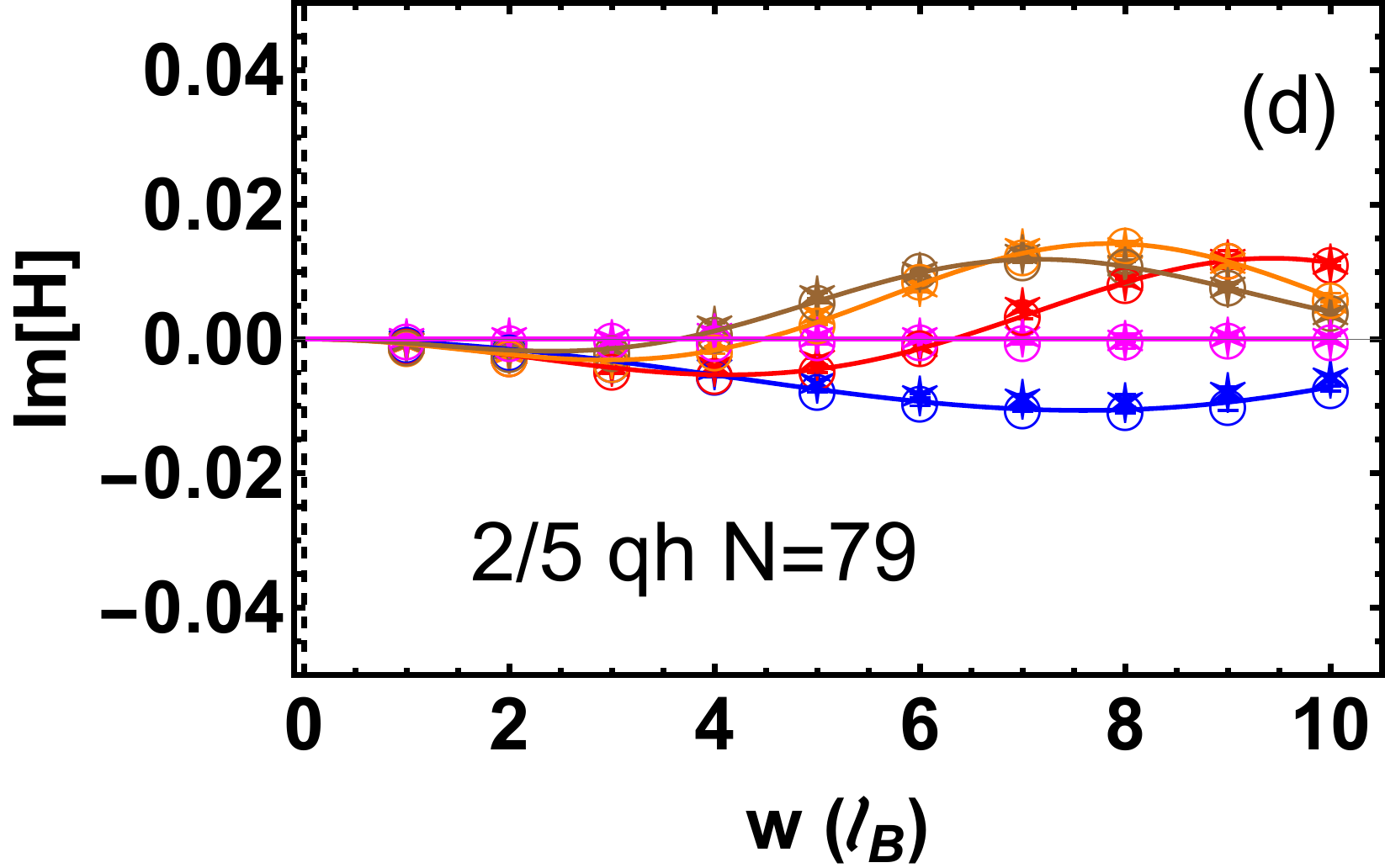} 
    \includegraphics[width=0.48\columnwidth]{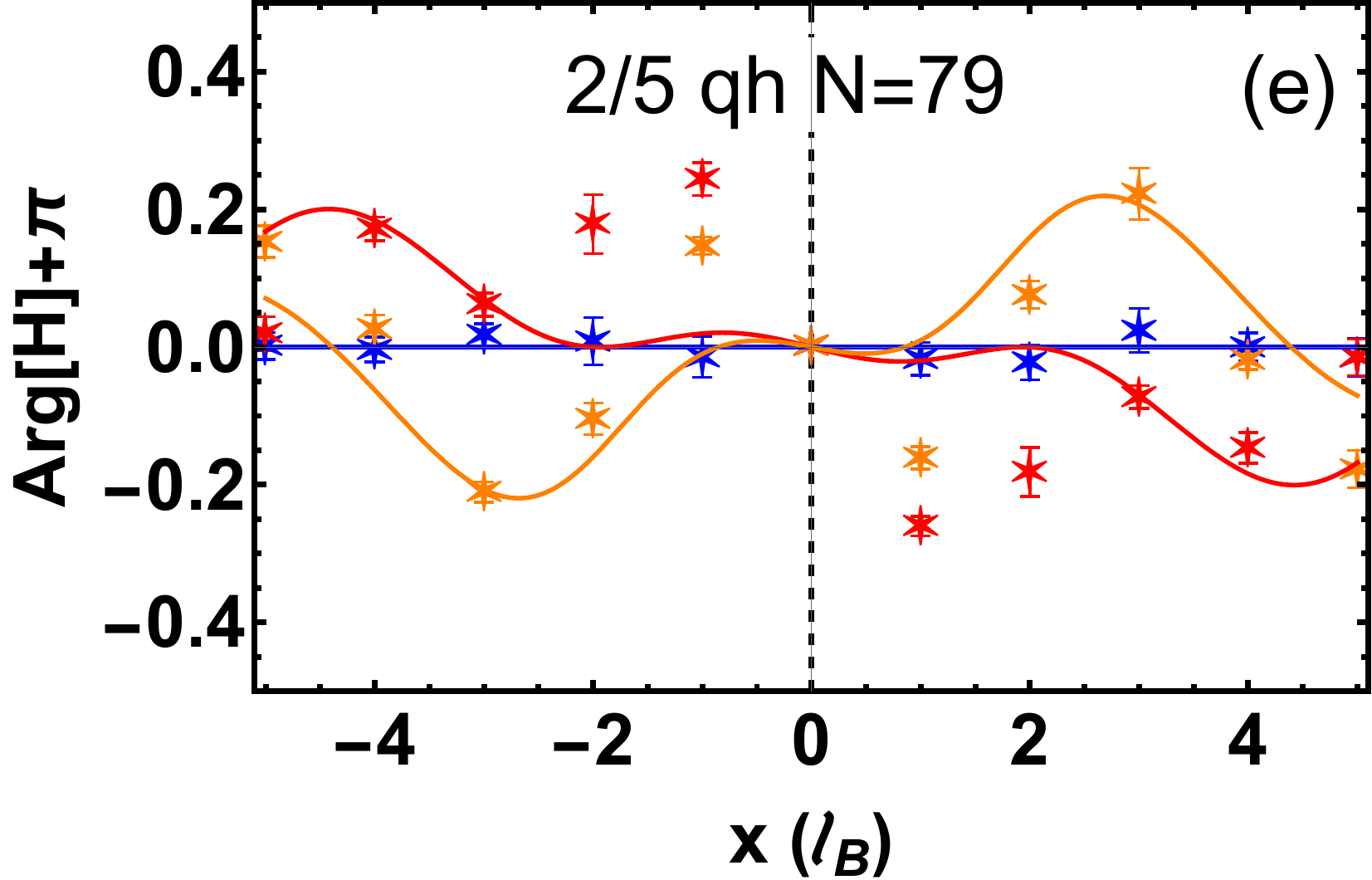} 
	\includegraphics[width=0.48\columnwidth]{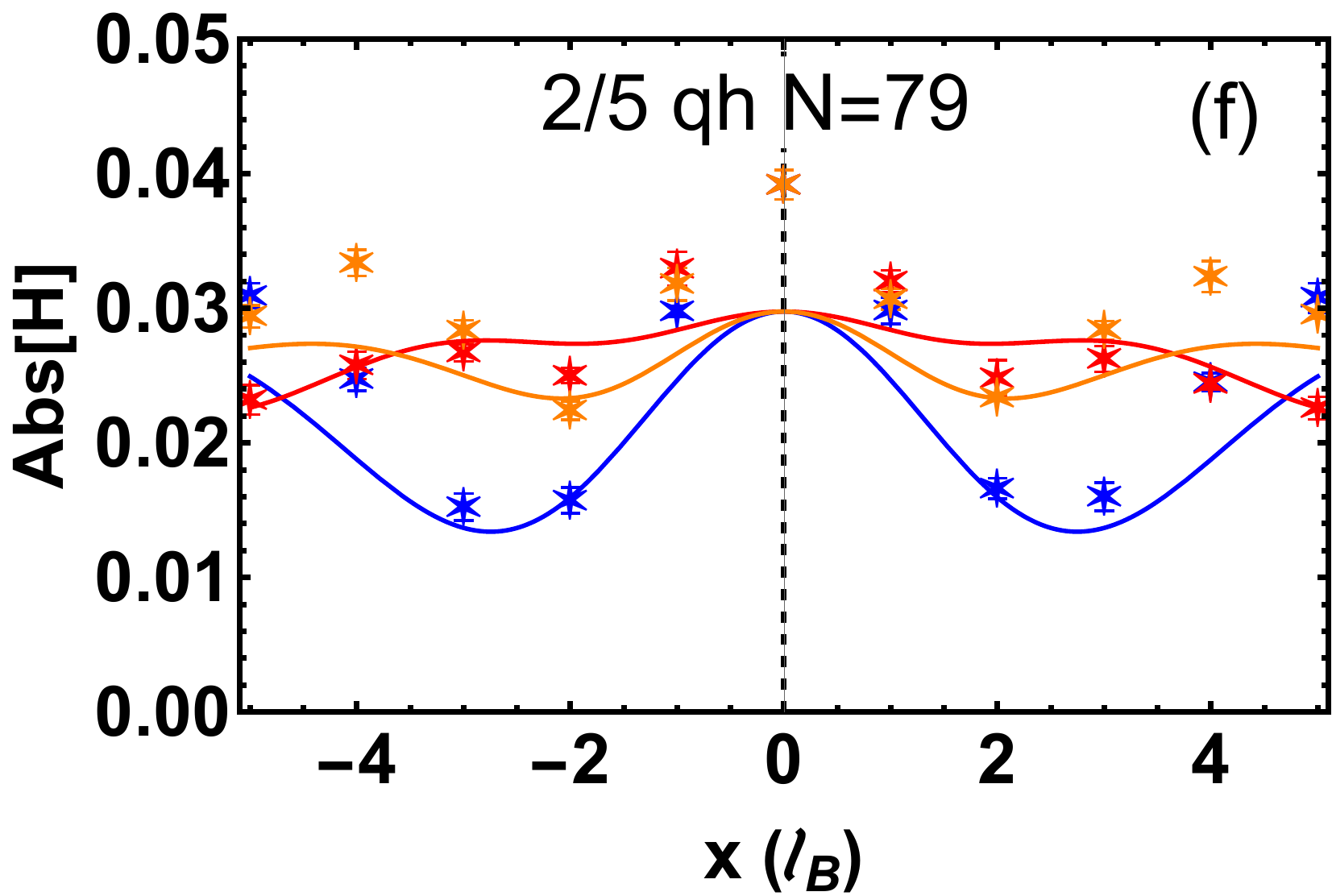} 
   \includegraphics[width=0.48\columnwidth]{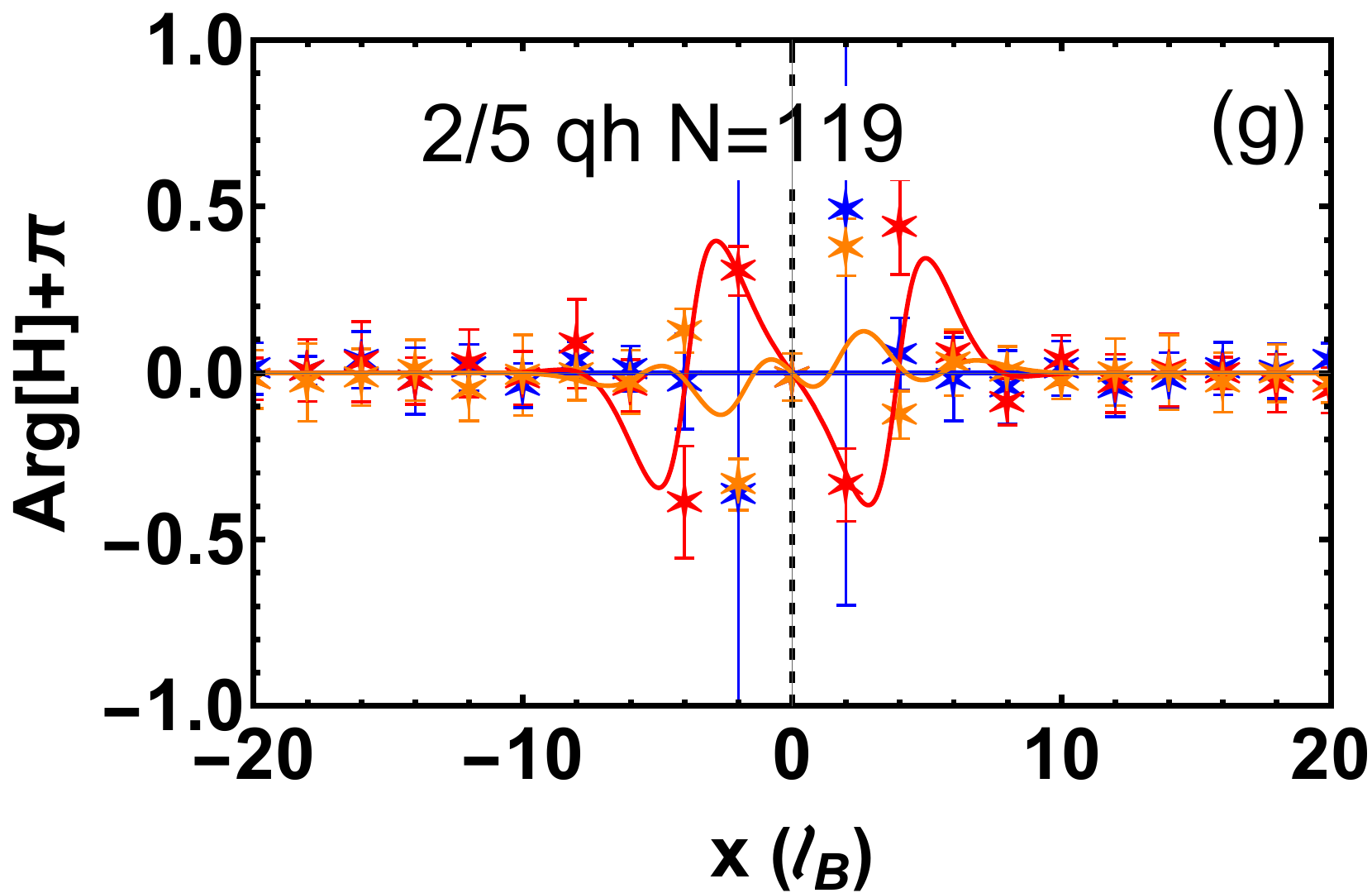} 
	\includegraphics[width=0.48\columnwidth]{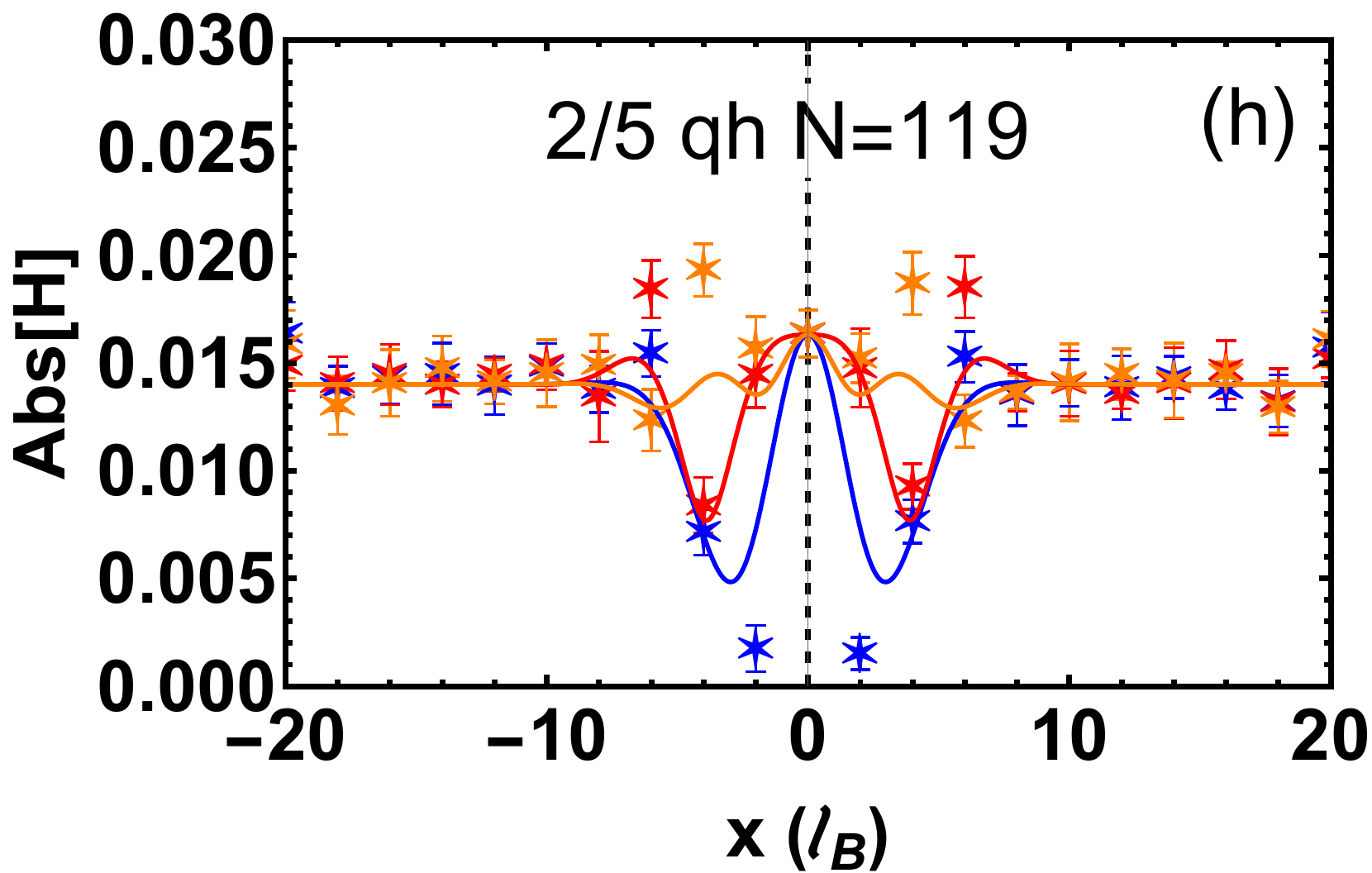} 
	\caption{This panel show various matrix elements for the quasihole of the $\nu=2/5$ state (stars) and compares them with the corresponding analytical results for an electron hole in the $n=1$ Landau level at an effective magnetic field (solid lines). Panels (a) and (b) show the real part and imaginary part of the overlap matrix elements $\langle \Psi^{{\rm qh}-2}_{{w\over 2}e^{i\theta}}|\Psi^{{\rm qh}-2}_{w\over 2}\rangle$, for $\theta=\pi/6$ (blue); $\theta=\pi/3$ (red); $\theta=\pi/2$ (orange); $\theta=\pi/2$ (brown); and $\theta=\pi$ (magenta).  Panels (c) and (d) show the real part and imaginary part of the tunneling matrix elements $\langle \Psi^{{\rm qh}-2}_{{w\over 2}e^{i\theta}}|\sum_i \delta\left(\vec{r}_i-{w\over 2}e^{i\theta}\right)|\Psi^{{\rm qh}-2}_{w\over 2}\rangle$ with the same color code.   Panel (e) - (h) show the phase and the modulus of the impurity assisted tunneling matrix element $\langle \Psi^{{\rm qh}-2}_{-{w\over 2}}|\sum_i \delta\left(\vec{r}_i-xe^{i\theta'}\right)|\Psi^{{\rm qh}-2}_{w\over 2}\rangle$, for $\theta'=0$ (blue);  $\theta'=\pi/4$ (orange); and $\theta'=\pi/2$ (red). Panel (e) and (f) correspond to $w=5$, and panel (g) and (h) correspond to $w=8$. All length are quoted in units of the magnetic length at $\nu=2/5$. The number of particles $N$ in the Monte Carlo calculation is shown on each panel; the results represent the thermodynamic limit. The larger uncertainty for small $x$ in panel (g) arises because the amplitude becomes very small here (see panel h).
 }
	\label{O25qh}
\end{figure}

\begin{figure}[]
	\includegraphics[width=0.48\columnwidth]{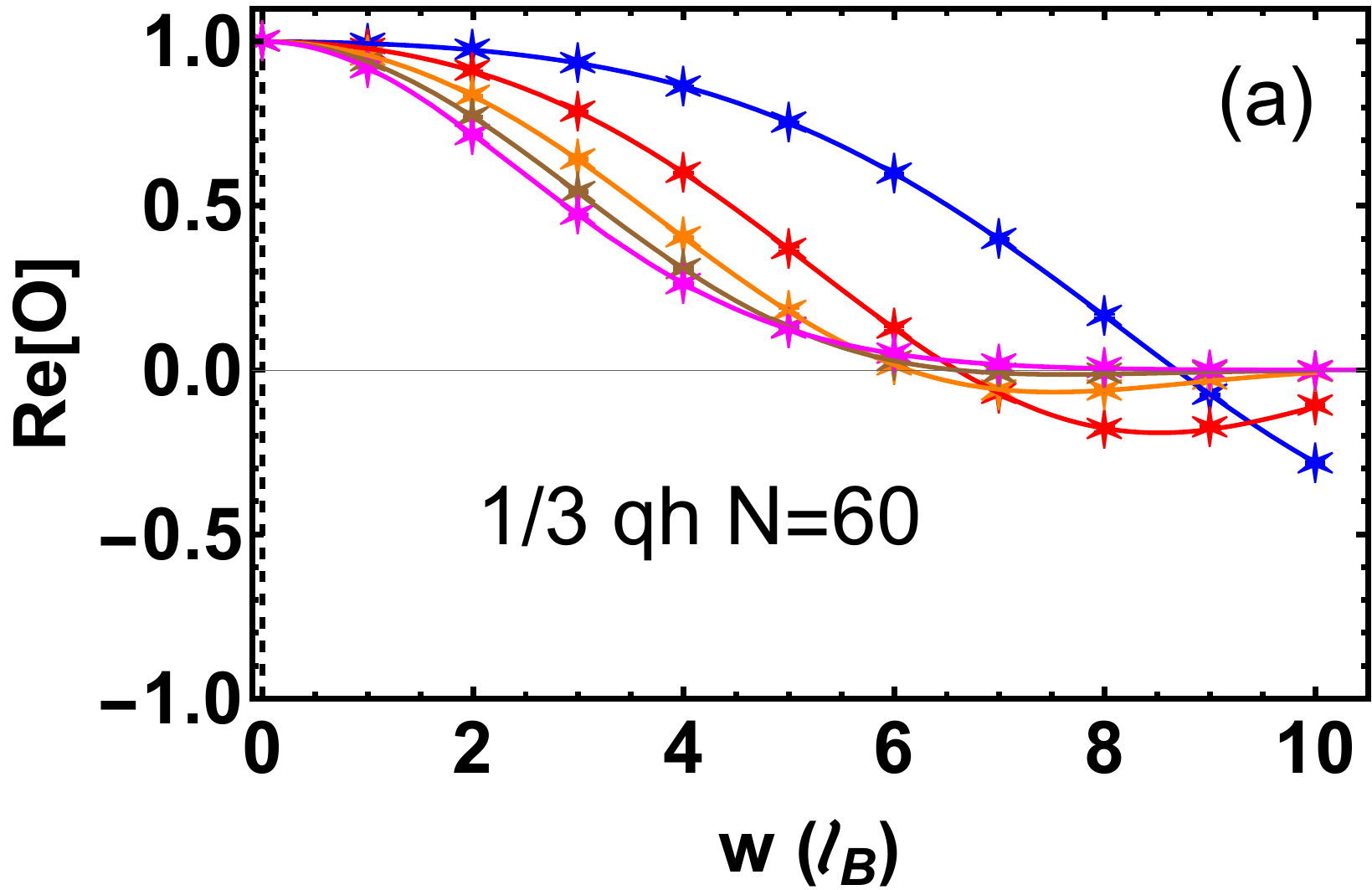} 
	\includegraphics[width=0.48\columnwidth]{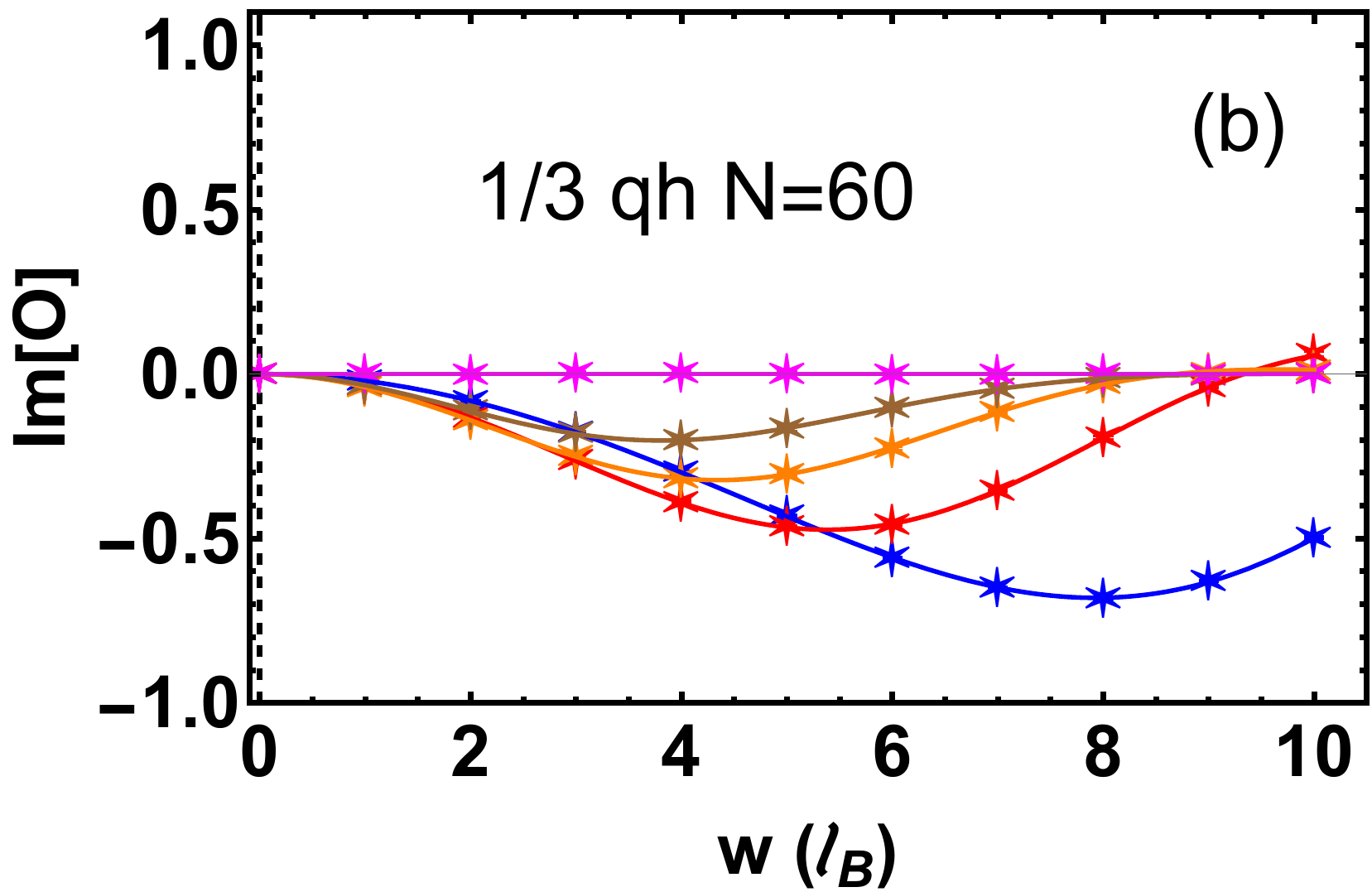} 
    \includegraphics[width=0.48\columnwidth]{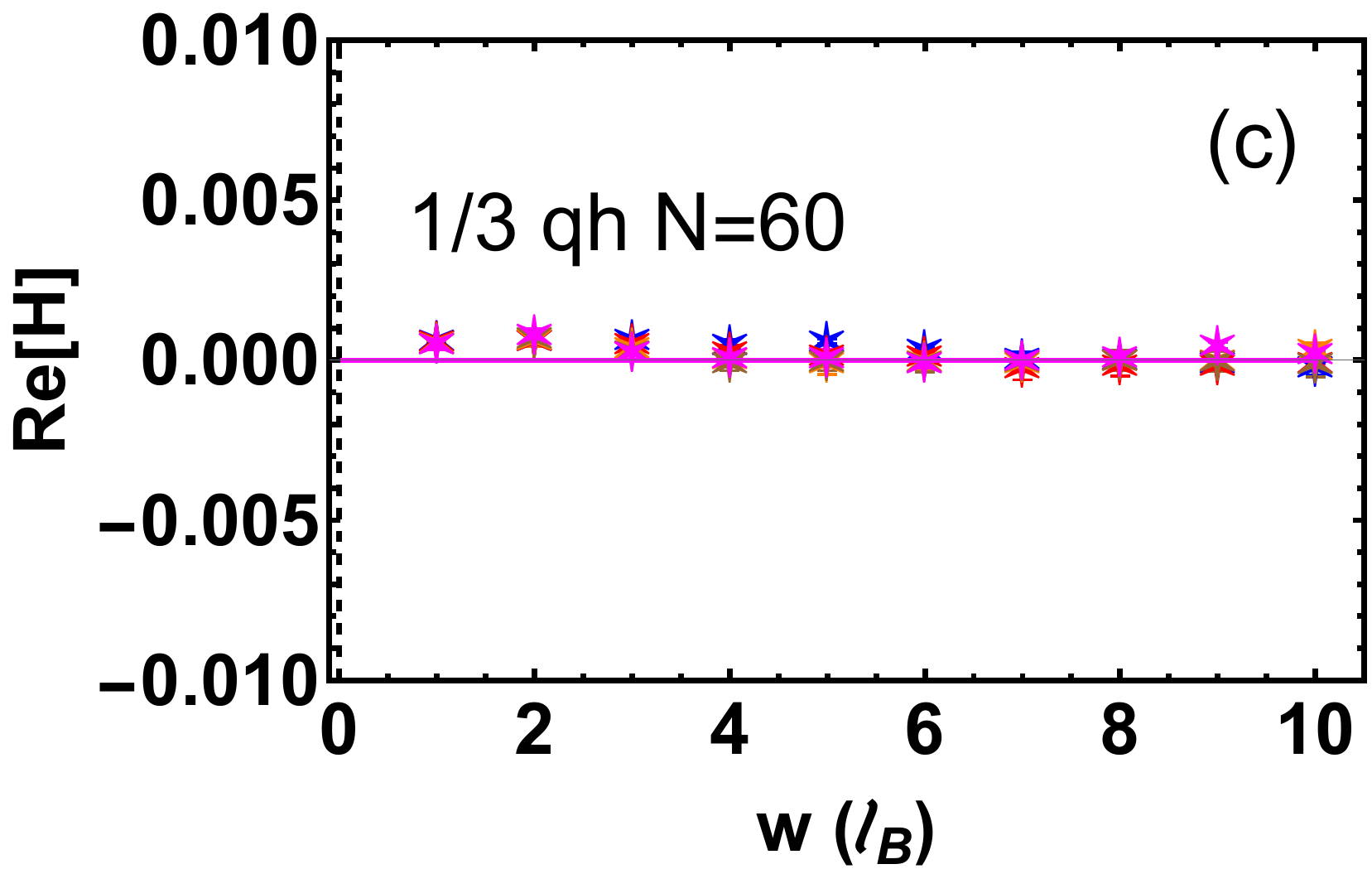} 
	\includegraphics[width=0.48\columnwidth]{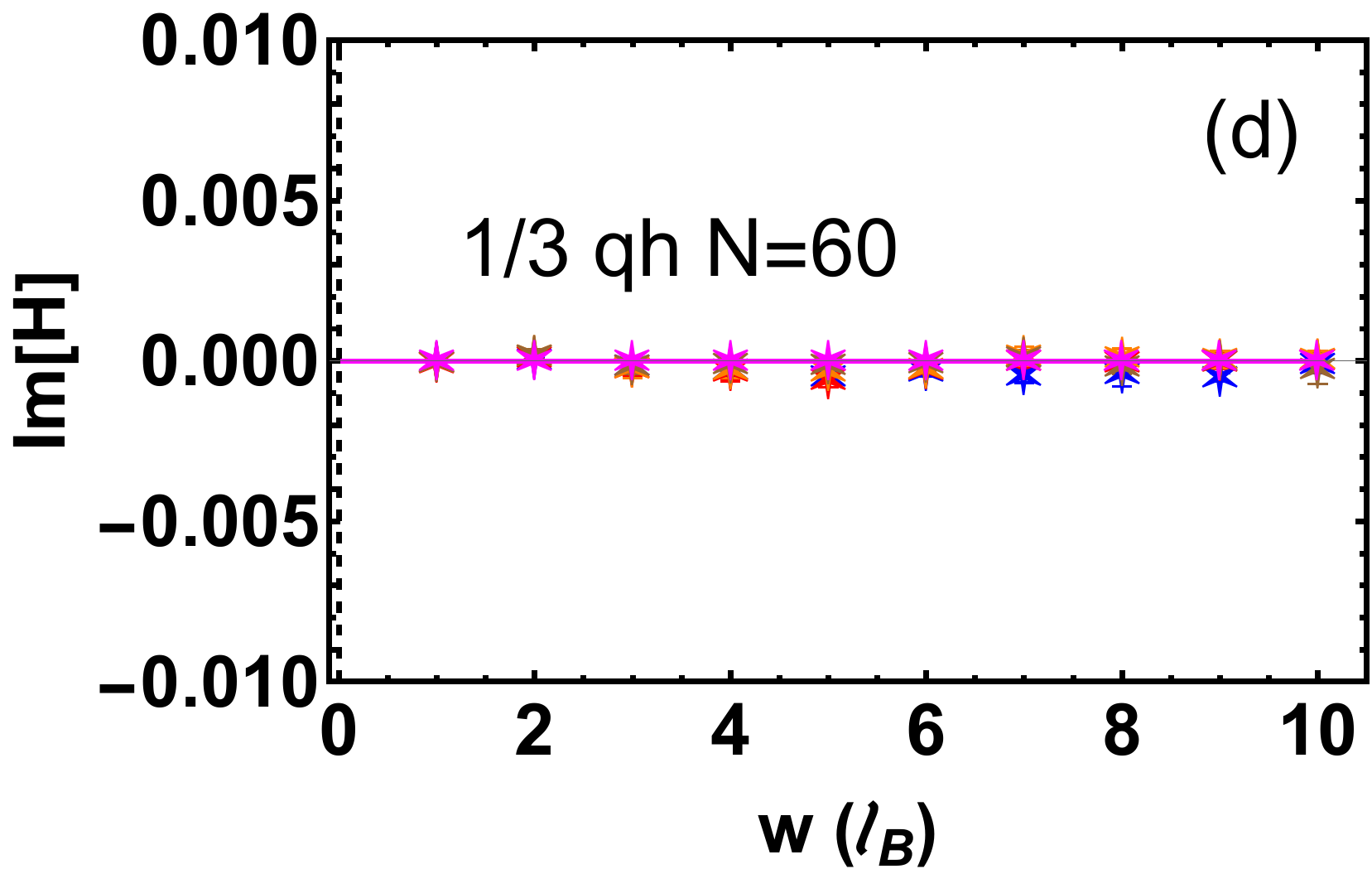} 
    \includegraphics[width=0.48\columnwidth]{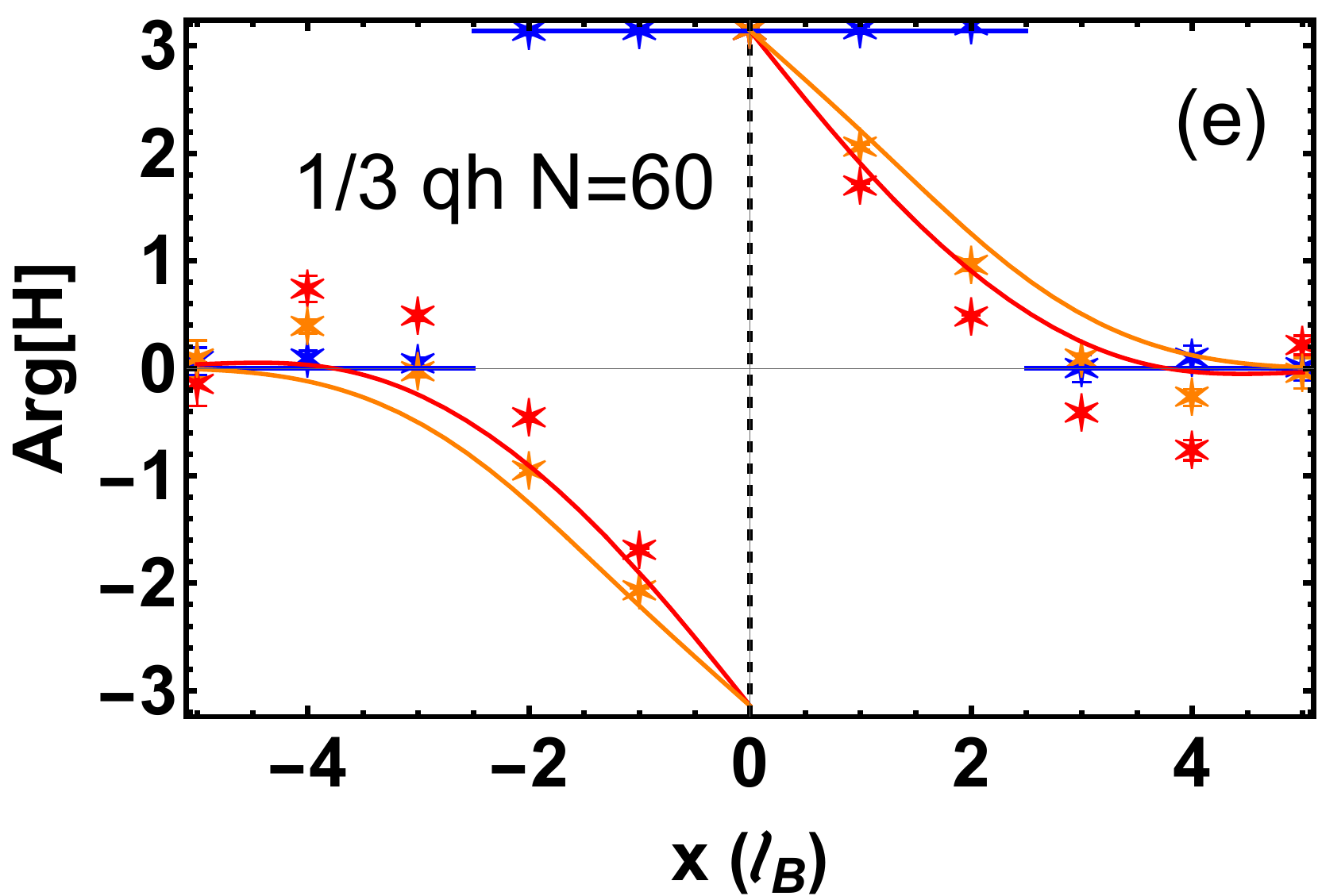} 
	\includegraphics[width=0.48\columnwidth]{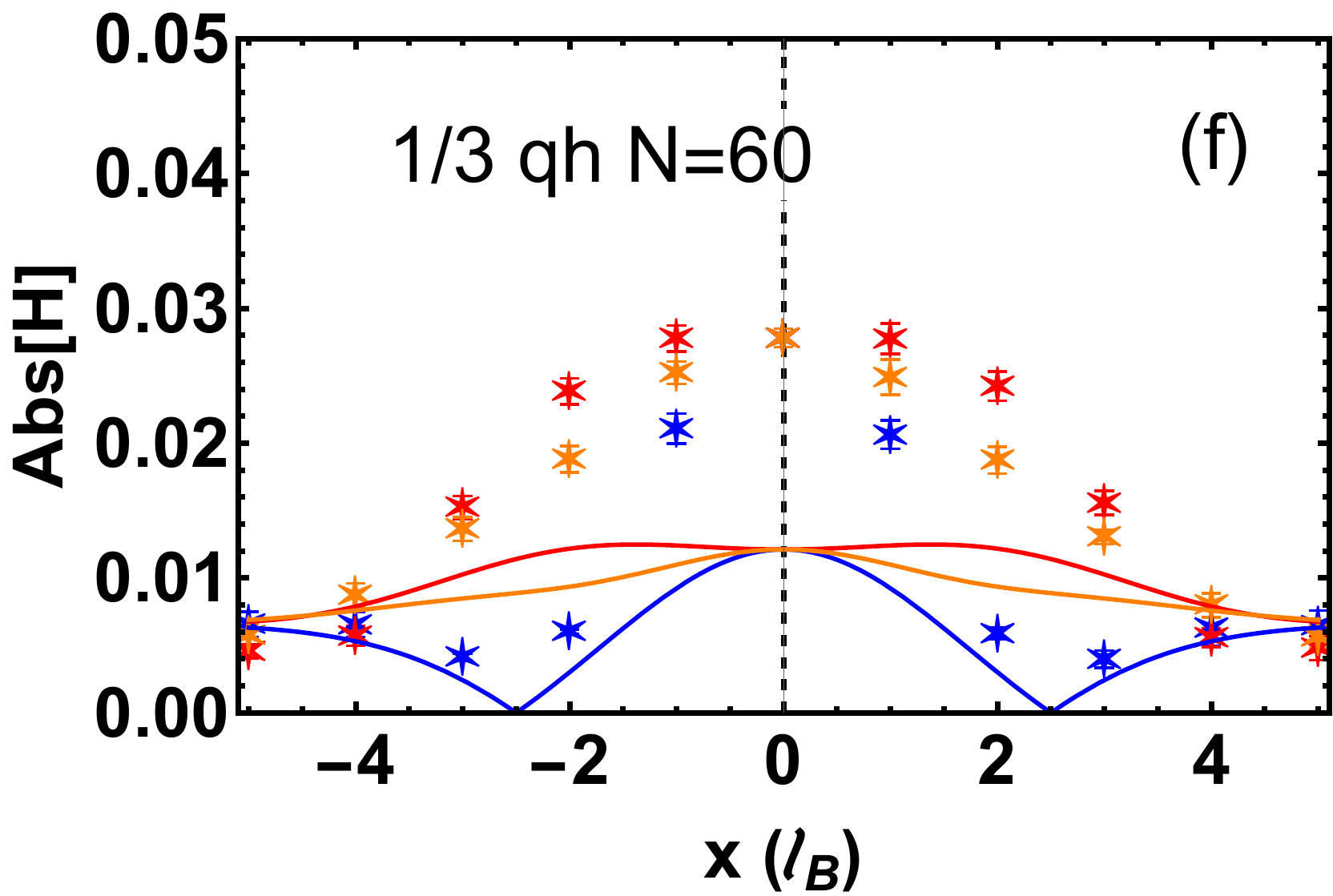} 
   \includegraphics[width=0.48\columnwidth]{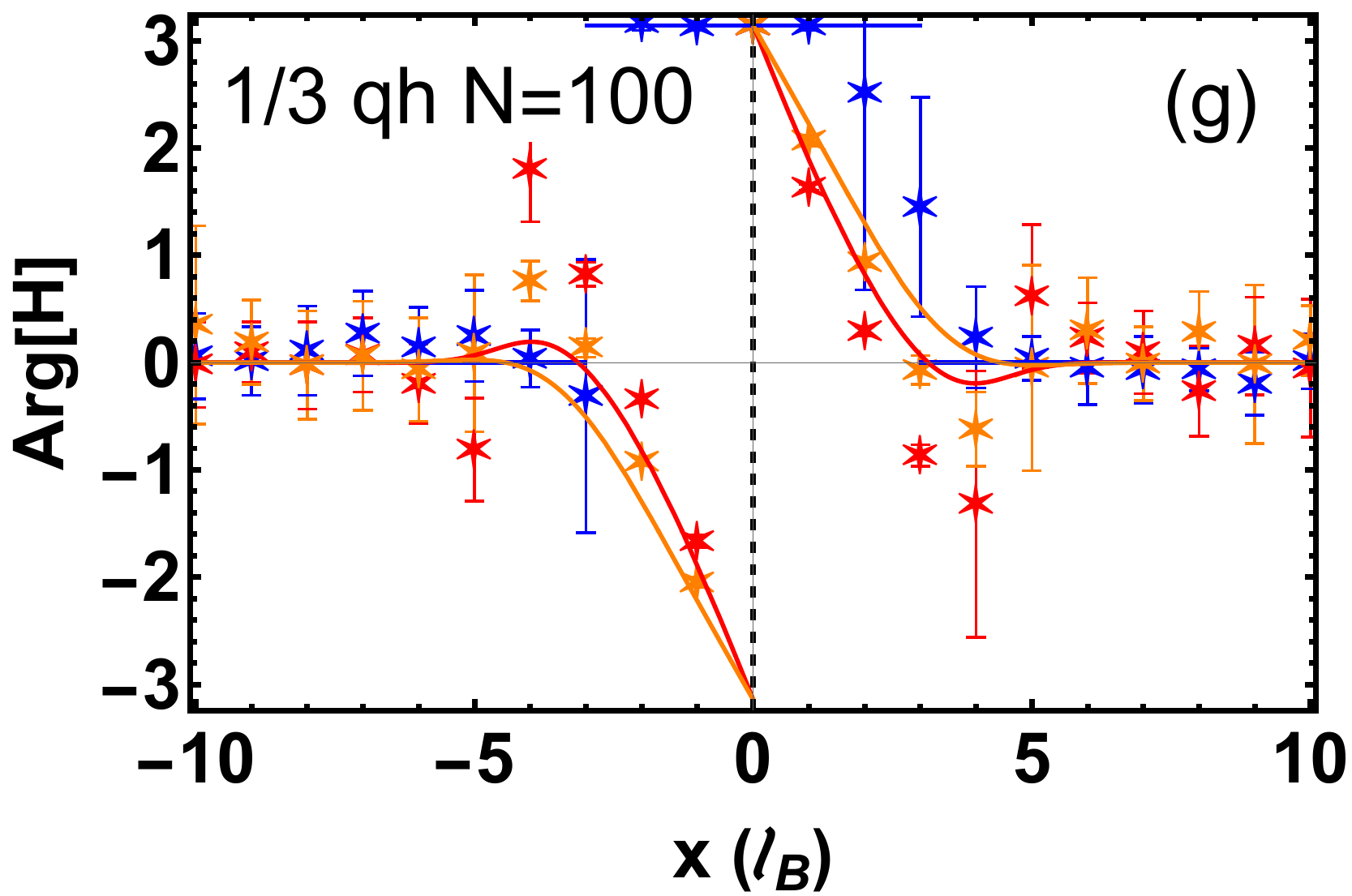} 
	\includegraphics[width=0.48\columnwidth]{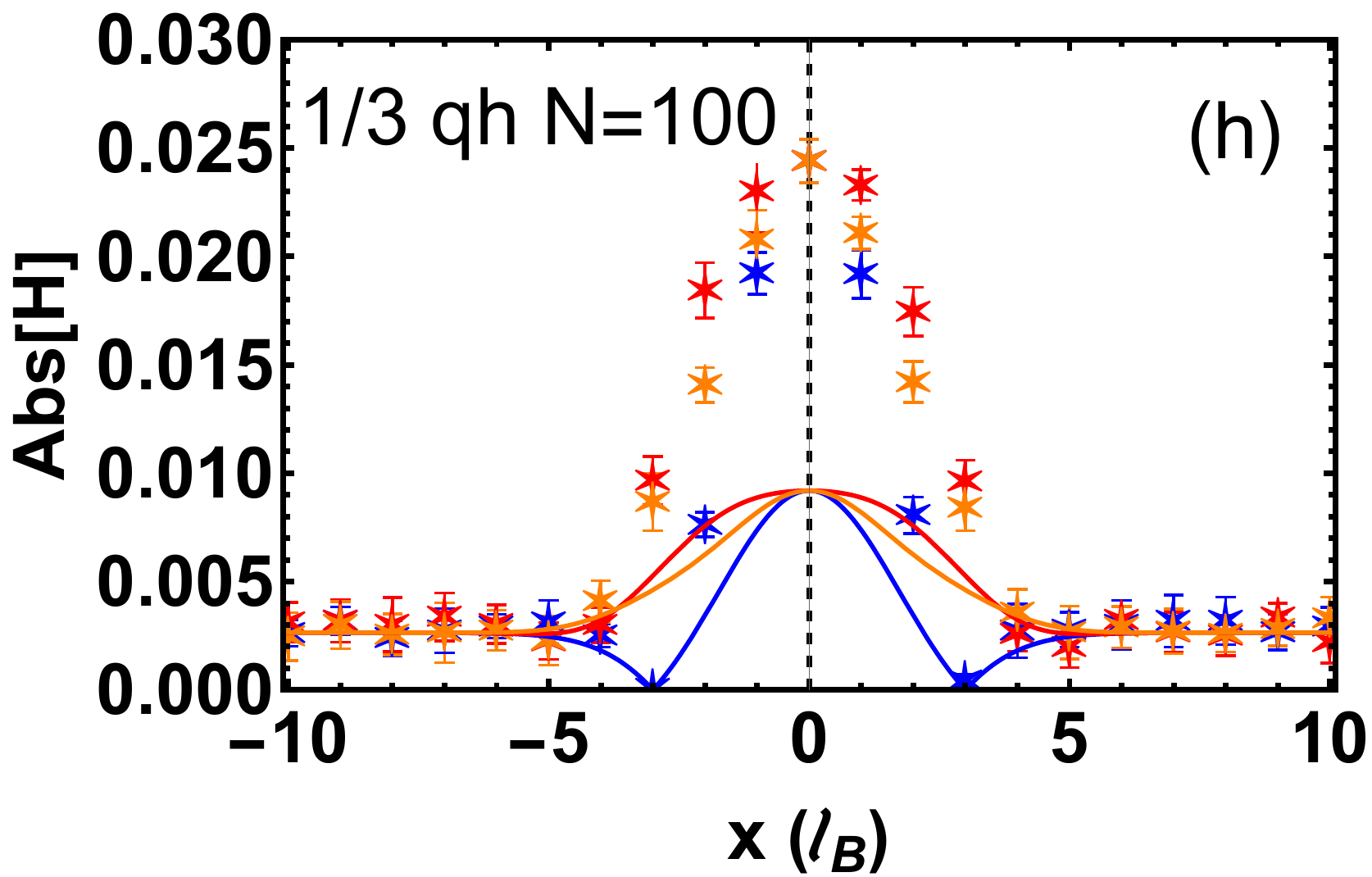} 
	\caption{This panel show various matrix elements for the quasihole of the $\nu=1/3$ state (stars) and compares them with the corresponding analytical results for an electron hole in the $n=0$ Landau level at an effective magnetic field (solid lines). Panels (a) and (b) show the real part and imaginary part of the overlap matrix elements $\langle \Psi^{{\rm qh}-1}_{{w\over 2}e^{i\theta}}|\Psi^{{\rm qh}-1}_{w\over 2}\rangle$, for $\theta=\pi/6$ (blue); $\theta=\pi/3$ (red); $\theta=\pi/2$ (orange); $\theta=\pi/2$ (brown); and $\theta=\pi$ (magenta).  Panels (c) and (d) show the real part and imaginary part of the tunneling matrix elements $\langle \Psi^{{\rm qh}-1}_{{w\over 2}e^{i\theta}}|\sum_i \delta\left(\vec{r}_i-{w\over 2}e^{i\theta}\right)|\Psi^{{\rm qh}-1}_{w\over 2}\rangle$ with the same color code. Panel (e) - (h) show the phase and the modulus of the impurity assisted tunneling matrix element $\langle \Psi^{{\rm qh}-1}_{-{w\over 2}}|\sum_i \delta\left(\vec{r}_i-xe^{i\theta'}\right)|\Psi^{{\rm qh}-1}_{w\over 2}\rangle$, for $\theta'=0$ (blue);  $\theta'=\pi/4$ (orange); and $\theta'=\pi/2$ (red). Panel (e) and (f) correspond to $w=5$, and panel (g) and (h) correspond to $w=6$. All length are quoted in units of the magnetic length at $\nu=1/3$. The number of particles $N$ in the Monte Carlo calculation is shown on each panel; the results represent the thermodynamic limit.
 }
	\label{O13qh}
\end{figure}

Figures~\ref{O25qh} and \ref{O13qh} depict the overlap and tunneling matrix elements for the quasihole of 2/5 and 1/3. he overlaps $O_{w_1w_2}$ and tunneling matrix elements $H^{{\rm qh}-n}_{w_1w_2,w_1}$ are well represented by the corresponding IQHE quantities $\mathcal{O}_{w_1w_2}$ and $\mathcal{H}^{{\rm qh}-n}_{w_1w_2,w_1}$. The correspondence between the general tunneling matrix element $H^{{\rm qh}-n}_{w_1,w_2;w}$ with the corresponding IQHE quantity $\mathcal{H}^{{\rm qh}-n}_{w_1,w_2;w}$, however, is not accurate, especially when the distance $|w_1-w_2|$ is small. 

\end{appendix}


\end{document}